\pdfoutput=1
\PassOptionsToPackage{table}{xcolor}

\documentclass[11pt]{article}

\usepackage{ACL2023}

\usepackage{times}
\usepackage{latexsym}

\usepackage{multirow}  
\usepackage{array}     
\usepackage{graphicx}  
\usepackage{amsmath}
\usepackage{colortbl}
\usepackage{longtable}
\usepackage{booktabs}
\usepackage{float} 
\usepackage[T1]{fontenc}
\usepackage{pdflscape}
\usepackage[breakable]{tcolorbox}

\usepackage{listings}

\usepackage{adjustbox}
\usepackage{multicol}
\usepackage{comment}

\usepackage{subcaption}
\usepackage{placeins}
\usepackage{capt-of}

\usepackage{ragged2e}
\usepackage{makecell}
\usepackage{nomencl}
\usepackage{fancyvrb}
\usepackage{framed}

\definecolor{lightgray}{gray}{0.9}

\usepackage[utf8]{inputenc}
\DeclareUnicodeCharacter{FEFF}{}
     
\usepackage{microtype}

\usepackage{inconsolata}

\title{An Automated LLM-based Pipeline for
Asset-Level Database Creation to Assess
Deforestation Impact
}

\author{
    Avanija Menon, Ovidiu Serban \\
    Imperial College London
}

\begin{document}
\maketitle
\begin{abstract}
The European Union Deforestation Regulation (EUDR) requires companies to prove their products do not contribute to deforestation, creating a critical demand for precise, asset-level environmental impact data. Current databases lack the necessary detail, relying heavily on broad financial metrics and manual data collection, which limits regulatory compliance and accurate environmental modeling. This study presents an automated, end-to-end data extraction pipeline that uses LLMs to create, clean, and validate structured databases, specifically targeting sectors with a high risk of deforestation. The pipeline introduces Instructional, Role-Based, Zero-Shot Chain-of-Thought (IRZ-CoT) prompting to enhance data extraction accuracy and a Retrieval-Augmented Validation (RAV) process that integrates real-time web searches for improved data reliability. Applied to SEC EDGAR filings in the Mining, Oil \& Gas, and Utilities sectors, the pipeline demonstrates significant improvements over traditional zero-shot prompting approaches, particularly in extraction accuracy and validation coverage. This work advances NLP-driven automation for regulatory compliance, CSR (Corporate Social Responsibility), and ESG, with broad sectoral applicability. 
\end{abstract}

\section{Introduction}
The European Union Deforestation Regulation (EUDR), effective December 30, 2025, mandates companies to verify that their products do not originate from recently deforested land \cite{EUDR}. With deforestation contributing 15\% to global CO$_2$
 emissions \cite{EnergyTransitions}, industries with high environmental risks require precise asset-level tracking. However, significant data gaps persist: 30\% of Forest 500 companies lack public deforestation commitments, and 85\% of financial institutions lack comprehensive deforestation policies \cite{forest5002024}. Creating a physical asset database is labour intensive \cite{GeoAssetProject2024}, costly, and inefficient, making regulatory compliance difficult and limiting researchers' ability to develop accurate environmental impact models.

Due to their substantial contributions to environmental degradation, we focus on three high-risk sectors—Mining, Oil \& Gas, and Utilities. Mining drives deforestation through surface extraction and infrastructure expansion, often leading to forest loss within a 50 km radius \cite{bradley2020miningforests}. Oil \& Gas exploration accelerates deforestation, particularly in biodiversity hotspots like the Amazon, where oil extraction disrupts ecosystems \cite{Finer2008, amazonwatch2016crude}. Utilities, especially hydroelectric projects, contribute to deforestation \cite{integritynext2024energyutilities} through extensive land clearing for dams and power infrastructure, with continued expansion affecting forested areas despite the shift to renewable energy \cite{pringle, IMPERIALE2023101468}.

\begin{figure}[H] 
    \centering
    \includegraphics[width=0.5\textwidth]{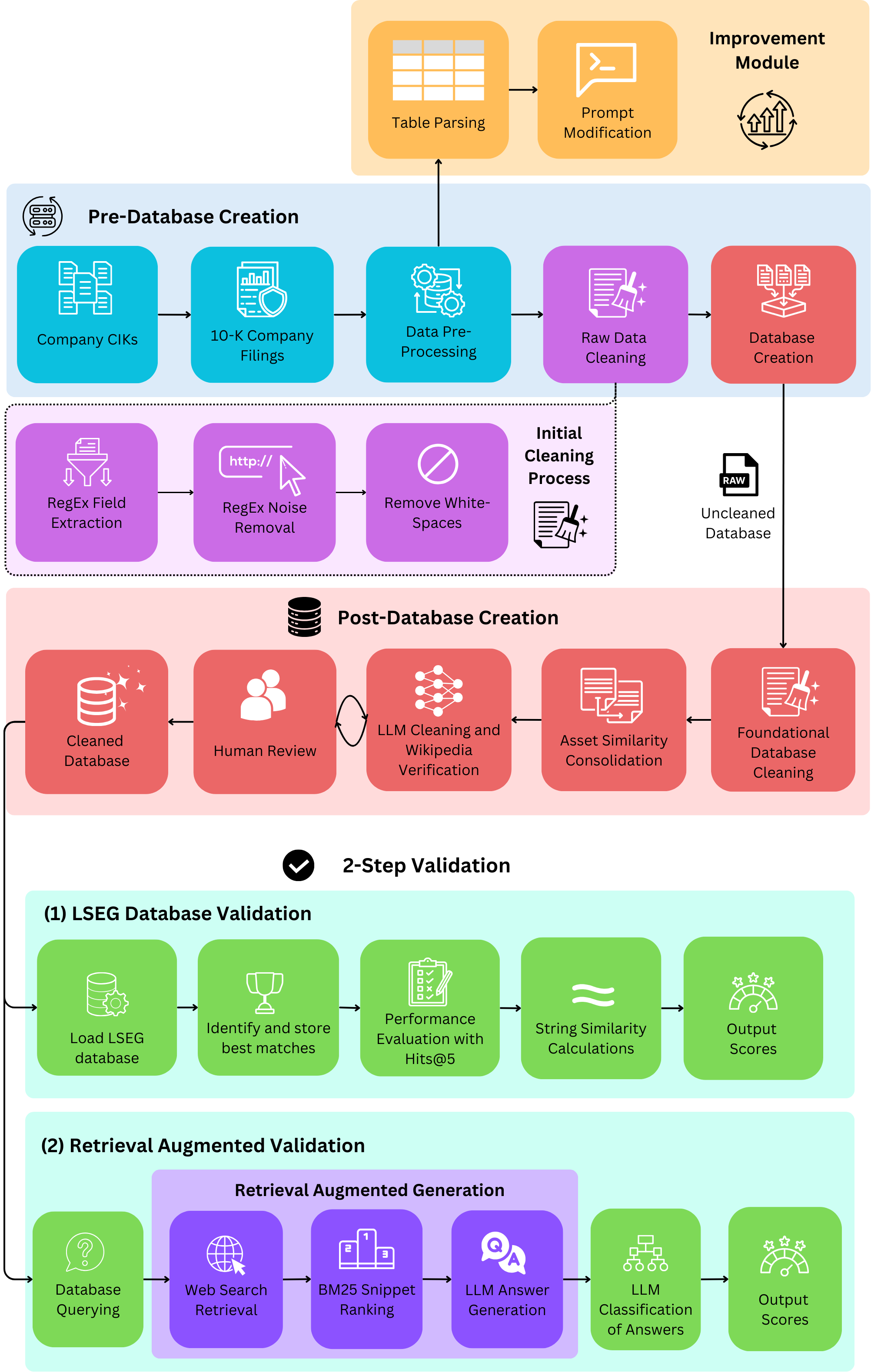}
    \caption{System design of end-to-end LLM-based pipeline designed to handle systematic data extraction, structured database creation, cleaning and validation, and the improvement module to increase validation coverage.}
    \label{fig:fullpipeline}
\end{figure}

Our research makes several key contributions: (1) We develop a novel LLM-based pipeline (Figure \ref{fig:fullpipeline}) that transforms unstructured SEC EDGAR filings into structured datasets, improving transparency in environmental monitoring. (2) We introduce Instructional, Role-Based, Zero-Shot Chain-of-Thought (IRZ-CoT) prompting, a technique that enhances the accuracy of entity extraction, particularly for complex asset-related information. (3) We conduct a comparative analysis of LLMs and a traditional Named Entity Recognition (NER) model, evaluating their effectiveness in domain-specific data extraction. (4) To ensure data integrity, we implement a three-step database cleaning process, which includes foundational standardisation, asset similarity consolidation using statistical methods, and LLM-assisted refinement. (5) We propose Retrieval-Augmented Validation (RAV), which integrates real-time web data to enhance dataset reliability and address gaps in existing databases. (6) Finally, the resulting datasets are visualised through company-specific dashboards, providing detailed insights into each company's database. 


This work advances NLP-driven environmental data automation, providing a scalable framework for regulatory compliance, sustainability analysis, and asset-based deforestation tracking. 

\section{Background and Related Work}

LLMs revolutionised entity and relation extraction, enabling zero-shot and few-shot learning. Structured prompting techniques, such as Pipeline Chain-of-Thought (Pipeline-COT), enhance accuracy by breaking tasks into reasoning steps \cite{pipelinecot}. ML and NLP techniques have been widely applied in healthcare, finance, and legal domains. Transformer-based models like LegalBERT \cite{legalbert}, BioBERT \cite{biobert}, and SciBERT \cite{scibert} improve clinical text analysis and regulatory compliance. However, fine-tuning remains computationally expensive, making zero-shot LLM approaches more practical. GPT-based models like GPT-NER incorporate self-verification to reduce hallucinations \cite{wang2023gptnernamedentityrecognition}, while ChatGPT and REBEL enable structured knowledge extraction \cite{trajanoska2023enhancingknowledgegraphconstruction}. This study builds on these advancements, introducing Instructional, Role-Based, Zero-Shot Chain-of-Thought (IRZ-CoT) prompting to enhance structured data extraction from SEC EDGAR filings.

Traditional SEC EDGAR processing relies on RegEx-based tools like LexNLP, which efficiently parse filings \cite{Bommarito2018LexNLP}. Prior keyword extraction and manual annotation work, such as the KPI-EDGAR dataset, remains labour-intensive and challenging to scale \cite{Deuer2022}.

Despite NLP advancements, limited work has been done on developing a fully automated pipeline that integrates data extraction, database creation, cleaning, and validation. This research bridges that gap by implementing an LLM-driven end-to-end pipeline, introducing Retrieval-Augmented Validation (RAV) to improve accuracy and robustness. By combining LLM-assisted extraction, structured prompts, and multi-step validation, this study delivers a scalable asset-tracking and environmental impact analysis solution, advancing AI-driven automation for regulatory data processing.

\section{Data Acquisition and Processing}

\subsection{Data Source}

This study uses publicly available SEC EDGAR 10-K filings from fifteen Mining, Oil \& Gas, and Utilities companies. These legally mandated reports provide standardised, reliable, and accurate data on company operations, finances, and environmental impact. Unlike 10-Q and 8-K reports, which offer limited asset details, 10-K filings comprehensively cover physical assets, expenditures, and disclosures. News and social media data were excluded due to bias, noise, and lack of granularity. SEC filings ensure factual accuracy, regulatory compliance, and ethical data sourcing, minimising legal and privacy concerns.

\subsection{Data Extraction}

We collected 10-K filings from 2022 to 2024 using the secEDGAR Python library \cite{sec-edgar}, which allows efficient bulk downloads based on company stock tickers and Central Index Keys (CIKs). This method streamlines data acquisition, eliminating the need for custom web scraping scripts while ensuring robust datasets across the selected sectors. The companies focused on are given in Table \ref{tab:companies} in Appendix \ref{stocktickers}.

The pre-processing workflow extracts metadata (company names, filing dates, form types, and content), cleans text using BeautifulSoup to remove HTML tags and irrelevant elements, and structures data into SQLite databases per company. This ensures efficient management, querying, and retention of meaningful content for analysis.

\section{Database Creation}

\subsection{Chunk-based Querying Technique}

We adopt a chunk-based querying technique to manage the extensive length of SEC EDGAR filings. This method involves splitting documents into 1024-token chunks with a 20-token overlap to maintain contextual continuity. Sentence-level splitting ensures semantic coherence, preventing the disruption of key information. Chunking optimises memory usage, enables parallel processing, and enhances entity recognition by allowing LLMs to focus on specific, contextually rich segments. This approach also facilitates error identification and correction, improving the efficiency and scalability of the data processing pipeline.

\subsection{Comparison of LLM and NER Outputs}
We compare the performance of 4-bit quantised Ollama instruct models, specifically Mistral-7B, Llama 3, and Gemma 2, against a traditional Named Entity Recognition (NER) model: dslim/bert-large-NER \cite{DBLP:journals/corr/abs-1810-04805, tjong-kim-sang-de-meulder-2003-introduction}. Instruct models, fine-tuned for instruction-based tasks, demonstrate superior contextual understanding and precise entity extraction \cite{chung2022scalinginstructionfinetunedlanguagemodels, hu2024finetuninglargelanguagemodels}, which could be used for structured documents like SEC filings. The use of 4-bit quantisation significantly reduces memory and computational requirements while maintaining performance, enabling efficient large-scale deployment without extensive hardware upgrades \cite{banner2019posttraining4bitquantizationconvolution, dettmers2023qloraefficientfinetuningquantized}. These models minimise irrelevant responses, ensuring more accurate asset identification. We convert the text data into embeddings using the SentenceTransformer model, specifically the paraphrase-MiniLM-L6-v2 variant \cite{reimers-2019-sentence-bert}. Gemma 2 consistently outperforms the NER model on cosine similarity metrics, achieving higher precision and recall, with the highest cosine similarity for both locations (0.7702) and organisations (0.7461), indicating strong alignment with ground truth data.

Error analysis reveals that LLMs are more effective in capturing nuanced entity relationships, while the NER model often fragments entities or misses domain-specific terms. Detailed performance metrics are provided in Table \ref{tab:llmvsner} in Appendix \ref{cosinesimilaritiesLLMNER}, where Gemma 2 outperforms both Mistral-7B and Llama 3.


As shown in Table \ref{tab:llm_analysis} in  Appendix \ref{erroranalysisLLMNER}, qualitative error analysis highlights common issues such as fragmented entity recognition in the NER model and occasional hallucination in LLM outputs. While Mistral-7B and Llama 3 struggled with consistency, Gemma 2 demonstrated more reliable extraction, particularly in complex texts.

\subsection{Ground Truth Creation}

We manually curated a ground truth dataset from 30 chunks of Alcoa Corporation's 2022 filings to evaluate extraction accuracy. This dataset includes detailed annotations of physical assets, their locations, ownership structures, and associated commodities. Manual annotation ensures high accuracy, providing a robust benchmark for model evaluation. While slightly labour-intensive, this process establishes a reliable foundation for assessing model performance. In future work, we recommend exploring automated ground truth generation using advanced models like GPT-4, which could enhance scalability and reduce annotation costs.

\subsection{LLM Selection}

We assessed multiple LLMs using evaluation metrics such as cosine and jaccard similarities, precision, recall, and F1 score. Gemma 2 emerged as the top performer, excelling in quantitative and qualitative analyses. Its superior performance is attributed to its ability to maintain semantic coherence and accurately extract domain-specific entities. As presented in Table \ref{tab:model_comparison}, Gemma 2 achieved the highest scores across all evaluation metrics. This performance consistency and efficient resource utilisation led to its selection for further experimentation within the data pipeline.
\begin{table}[H]
    \centering
    \renewcommand{\arraystretch}{1.2} 
    \setlength{\tabcolsep}{4pt} 
    \scriptsize 
    \begin{tabular}{|l|c|c|c|c|c|}
        \hline
        \multirow{2}{*}{\textbf{Model}} & \multicolumn{2}{c|}{\textbf{Similarity Metrics}} & \multicolumn{3}{c|}{\textbf{Evaluation Metrics}} \\
        \cline{2-6}
        & \textbf{Cosine} & \textbf{Jaccard} & \textbf{Precision} & \textbf{Recall} & \textbf{F1 Score} \\
        \hline
        Mistral-7B & 0.64 & 0.41 & 0.54 & 0.58 & 0.60 \\
        Llama 3 & 0.64 & 0.43 & 0.58 & 0.64 & 0.59 \\
        Gemma 2 & 0.68 & 0.44 & 0.63 & 0.62 & 0.60 \\
        \hline
    \end{tabular}
    \caption{Performance comparison of Mistral-7B, Llama 3, and Gemma 2 across five metrics.}
    \label{tab:model_comparison}
\end{table}

\subsection{Prompt Engineering}

Prompt engineering plays a crucial role in optimising data extraction. We developed the Instructional, Role-Based, Zero-Shot Chain-of-Thought (IRZ-CoT) prompting technique through iterative refinement. This method improves extraction accuracy by providing LLMs with domain-specific instructions, structured reasoning steps, and role-based guidance. IRZ-CoT reduces hallucination, enhances the extraction of complex attributes, and minimises the need for extensive post-processing.

Common issues encountered with different prompting techniques revealed key challenges, such as hallucination in one-shot and few-shot methods, incorrect classification in zero-shot, and verbosity in generated knowledge prompting. Specifically, zero-shot prompting often led to the misclassification of financial terms as physical assets, while few-shot techniques introduced hallucinated entities. Role-based and instructional prompting significantly improved specificity and reduced errors, but IRZ-CoT demonstrated the best balance between accuracy and efficiency.

\begin{figure}[H] 
    \centering
    \includegraphics[width=0.5\textwidth]{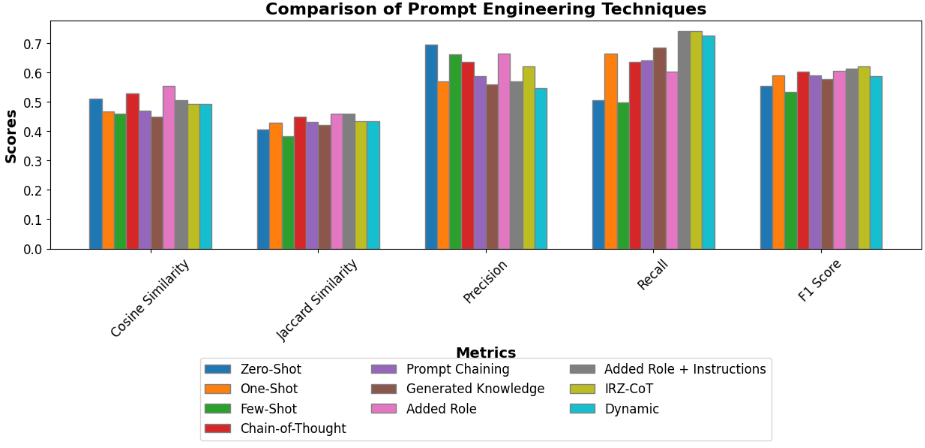} 
    \caption{ Comparison of different prompt engineering techniques across various evaluation metrics}
    \label{fig:allprompteng}
\end{figure}

Performance metrics for prompt engineering techniques, illustrated in Figure \ref{fig:allprompteng}, show that IRZ-CoT achieved the highest scores in precision and recall. Additionally, Figure \ref{fig:prompttime} in Appendix \ref{prompttime} highlights IRZ-CoT's computational efficiency, requiring significantly less processing time than methods like generated knowledge prompting.

\subsection{Experimental Evaluation of LLM Ensemble Methods}

We evaluated three LLM ensemble methods to enhance robustness: Ensemble Averaging with Majority Voting (EAMV), Weighted Majority Voting Ensemble (WMVE), and Stacking Ensemble with Meta-Learning (SEML). EAMV improves stability by aggregating predictions from multiple LLMs and selecting the most common output, reducing variance. WMVE assigns higher weights to models with superior performance, prioritising predictions from more accurate models, particularly favouring Gemma 2. SEML utilises a meta-learner—logistic regression—to combine outputs from different LLMs, optimising predictive accuracy and achieving the highest F1-score (Figure \ref{fig:ensembleresults} in Appendix \ref{ensembleresults}). However, SEML significantly increased processing time—nearly 20-fold compared to single-model approaches (Figure \ref{fig:ensembletimes} in Appendix \ref{ensembletimes}). Due to computational constraints, we selected the more efficient IRZ-CoT approach with Gemma 2 as the primary model for the final pipeline.

\section{Database Cleaning}

\subsection{Foundational Data Cleaning and Standardization}

The first phase focuses on refining raw data to establish a solid foundation for further processing. We use regular expression (RegEx) patterns to extract key entity data, including asset types, locations, ownership details, and commodities. This automated approach ensures consistent data extraction from large volumes of text. Post-extraction, we remove extraneous characters, such as surplus quotes and brackets, to prevent data distortion.

Duplicates are identified and consolidated, with corresponding information merged into single records. For example, multiple entries for an oil well in different locations are grouped, reducing redundancy. Ownership data is standardised by normalising company names (e.g., consolidating "NEM," "Newmont," and "Newmont Corporation"). At the same time, geographic terms are unified (e.g., "United States of America," "US," and "U.S.A." standardised to "USA"). We also refine the `location' column, extracting country names into a new `Countries' column to support consistent geographic analysis. Finally, rows with empty `physical asset' entries are removed to maintain database relevance.

\subsection{Asset Similarity Consolidation}

After initial cleaning, we address semantic similarities among physical asset entries. To consolidate such similarities, we use TF-IDF (Term Frequency-Inverse Document Frequency) vectorisation and cosine similarity.

TF-IDF quantifies the relevance of words within documents, and cosine similarity identifies semantic overlap. We set a similarity threshold of 0.5; entries meeting or exceeding this threshold are grouped and merged, preserving unique information while eliminating redundancy. This method is computationally efficient and effective for identifying similar assets, although it has limitations, such as sensitivity to synonyms. Despite these, TF-IDF and cosine similarity offer a pragmatic balance of accuracy and efficiency for large-scale datasets.

\subsection{LLM-Assisted Database Cleaning}

The final cleaning phase leverages the capabilities of Gemma 2 to address issues beyond the reach of traditional methods, as the previous steps still outputted unnormalised information amongst other issues. Using a domain-specific prompt, the LLM performs tasks such as converting chemical symbols (e.g., "Au" to "Gold"), standardising text, eliminating redundant punctuation, and verifying locations against Wikipedia.

This iterative process involves LLM-driven cleaning followed by human review. Any inconsistencies trigger prompt adjustments, enhancing the LLM's performance in subsequent iterations. The LLM also identifies countries from location data when not explicitly stated, verified through cross-referencing with Wikipedia to ensure accuracy.

By automating complex tasks and reducing manual effort, LLM-assisted cleaning improves data quality, consistency, and scalability, making it an effective strategy for managing large datasets. 

\section{Database Validation}

\subsection{Validation with LSEG Databases}

We validated our databases against established LSEG Workspace databases \cite{LSEG_Workspace}, focusing on the ‘Mines’, ‘Oil Refineries’, and ‘Power Generation’ datasets. This validation process involved data preprocessing, where we standardised text to lowercase and filtered irrelevant entries, such as excluding closed or abandoned assets. This step ensured uniformity and minimised discrepancies related to case sensitivity. Subsequently, we used the \texttt{rapidfuzz} library to find similar entries between our database entries and LSEG data. A similarity threshold of 0.6 was applied to identify potential matches, which helped us find the best match from the list of matches. We then used the Hits@5 metric to determine how frequently correct matches appeared within the top five candidates for each attribute (physical asset, ownership, commodity, and country). The Hits@5 score measures the consistency of our matching algorithm by averaging successful matches across all entities, assessing performance beyond the top result.
Identified matches are then validated using five more metrics (Partial Match Score (Partial Ratio), Jaccard Similarity, Cosine Similarity, Dice-Sørensen Similarity Coefficient and Normalised Levenshtein Distance), comparing entity similarities with the LSEG database. Detailed similarity scores across physical asset name, ownership, commodity, and country are averaged into an overall attribute similarity score, quantifying dataset alignment. This validation ensures the reliability of our matching algorithm.

\subsection{Retrieval-Augmented Validation (RAV)}
LSEG Workspace databases lack comprehensive data for complete validation, necessitating an additional verification layer to ensure completeness and accuracy. To address these gaps, we developed Retrieval-Augmented Validation (RAV).

RAV integrates real-time web search capabilities using the Google Custom Search Engine (CSE) API  \cite{google_custom_search_api} to retrieve current information on physical assets. The retrieved snippets are ranked using the BM25 algorithm, which prioritises documents based on relevance, incorporating term frequency and document length normalisation. This ensures that the most pertinent information is considered for validation purposes.

RAV uses a dual-LLM framework where Llama 3 generates web-based answers, and Gemma 2 is tasked with classifying these answers strictly against the database entries. This separation mitigates potential biases arising from using a single model for generation and evaluation. Llama 3 efficiently retrieves concise, relevant information from web sources, while Gemma 2 assesses the similarity between this information and the existing database entries. The LLM-assisted validation relies on a binary classification approach where Gemma 2 outputs a `yes' if the web-derived and database information are similar and a `no' otherwise. This stringent evaluation ensures high reliability, reducing the risk of false positives in validation.

Contrary to complex instructional prompts used in earlier phases, we discovered that simple prompts significantly improved LLM classification accuracy. Initially, using detailed prompts resulted in low similarity scores, averaging around 0.15, with frequent misclassifications. After simplifying the prompts to a single-line instruction asking the LLM to classify answers as similar or dissimilar (see Appendix \ref{sec:promptlibrary}), we observed a substantial improvement, with scores increasing by approximately 0.28. This reduction in cognitive load enhanced the model’s ability to determine similarities accurately. However, some misclassifications remain, mainly when subtle semantic differences exist between the database entries and web-sourced information.

RAV automates asset validation by integrating web data with traditional databases, enhancing reliability for downstream analysis. While advanced RAG methods like FLARE \cite{jiang2023active} offer sophisticated retrieval, their complexity and resource demands outweigh the benefits for this project. Our BM25-based RAV remains practical and effective, with potential for future refinement.

\section{Results}

\subsection{LSEG Database Validation Results}

\begin{figure}[H] 
    \centering
    \includegraphics[width=0.5\textwidth]{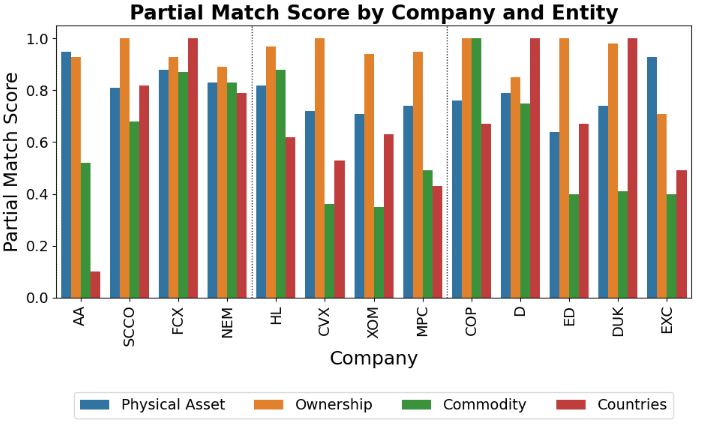} 
    \caption{Partial match scores from the LSEG database validation. The dotted lines separate the sectors, where the sectors are mining,
oil \& gas, and utilities, respectively.}
    \label{fig:partialmatch}
\end{figure}

We validated our databases against LSEG Workspace datasets, including `Mines,' `Oil Refineries,' and `Power Generation,' using six similarity metrics: Partial Match Score, Jaccard Similarity, Cosine Similarity, Dice-Sørensen Similarity Coefficient, Normalised Levenshtein Distance, and Hits@5. These metrics assessed alignment across physical assets, ownership, commodities, and country data. Figure \ref{fig:partialmatch} shows the partial match scores. The full results are shown in Figure \ref{fig:LSEGvalresults} in Appendix \ref{LSEGvalresults}.

The mining sector demonstrated strong alignment, with high partial match scores for companies such as AA (0.95), FCX (0.88), and NEM (0.83), reflecting data consistency. However, the oil \& gas sector showed mixed results, where CVX (0.72) and MPC (0.74) achieved moderate alignment, but XOM exhibited lower scores (Jaccard Similarity: 0.24), suggesting inconsistencies in asset classification. Utilities displayed moderate alignment, with EXC and D achieving partial match scores of 0.93 and 0.79, respectively. Ownership data varied significantly, with CVX achieving near-perfect alignment (1.00), while XOM had lower similarity (Jaccard Similarity: 0.43), likely due to differences in how joint ventures and subsidiaries were recorded. Commodity data showed the most significant discrepancies, with many companies, such as AA, registering low Jaccard and Cosine Similarities (0.00), possibly due to differences in classifying primary and secondary commodities. In contrast, country data was generally consistent, with companies like FCX achieving perfect alignment (Partial Match Score: 1.00), though some discrepancies were observed in SCCO (Jaccard Similarity: 0.67).  

The error analysis revealed key challenges. Ownership discrepancies arose due to variations in recording structures, where our databases captured joint ventures while LSEG focused on primary controlling entities. Standardising ownership classification could improve future alignment. Commodity misalignments resulted from differences in listing primary versus secondary commodities, suggesting a need to refine entity extraction prompts and separate commodity categories. Minor inconsistencies in country data, such as listing "USA" versus "California," highlight the importance of hierarchical structuring with separate fields for city, region, and country to enhance accuracy.

\subsection{Coverage Calculation}
We evaluate database coverage by measuring the proportion of physical assets and their attributes in our constructed database that match those in LSEG. This assessment ensures comprehensiveness, usability, and accuracy while identifying areas for improvement in our extraction pipeline. Coverage is computed as \( \text{Coverage Score} = \left( \frac{N_m}{N_L} \right) \times 100 \), where \( N_m \) represents the number of matched physical assets between our database and LSEG, and \( N_L \) is the total number of physical assets in LSEG. The computed coverage scores, shown in Table \ref{tab:coverage_table} in Appendix \ref{coverageresults}, indicate that the mining sector has better coverage than oil \& gas and utilities.

Manual inspection of SEC EDGAR filings reveals that lower coverage in oil \& gas and utilities stems from improper table parsing, as many assets are listed in tabular formats rather than continuous text. To address this, we integrated a table parsing module using LlamaIndex, which processes HTML tables as structured data instead of narrative text. This significantly improved extraction accuracy, particularly in oil \& gas, where assets were previously missed. Figure \ref{fig:coveragebeforeaftertableparse} in Appendix \ref{coveragetableparse} demonstrates this enhancement.

\subsection{Retrieval-Augmented Validation (RAV) Results}

Table \ref{tab:RAVresults} in Appendix \ref{RAVresults} presents the results of Retrieval-Augmented Validation (RAV), comparing database responses with real-time web data. Similarity scores range from 0.31 to 0.57, indicating moderate alignment, with the Oil \& Gas sector performing slightly better due to more transparent regulatory disclosures. Notably, OXY is absent from Table \ref{tab:RAVresults} since it only contains unnamed assets, which RAV cannot validate.

Mining sector scores hover around 0.4, suggesting uniform discrepancies, likely due to outdated or incomplete records. The Oil \& Gas sector shows slightly higher alignment, with companies like MPC and COP exceeding 0.5, possibly due to stringent regulatory reporting. However, frequent asset transfers contribute to inconsistencies. The Utilities sector exhibits the widest score range, from 0.31 (EXC) to 0.57 (NEE), reflecting differences in data transparency. NEE’s higher score suggests more consistent asset records, likely due to better data management.

\subsubsection{Error Analysis}

Ownership mismatches arose from differing data granularity. Our database captured joint ventures and minority stakeholders, whereas web sources listed only primary entities, leading to unfair scores of 0. A weighted scoring system could better account for partial matches.

Location mismatches often resulted from implicit references in web snippets. For instance, the Bath County Power Station was correctly labeled as USA in our database, but the web snippet lacked an explicit country mention, receiving a score of 0. Similarly, Chino Mine was recorded as USA, while web sources specified New Mexico, USA. A hierarchical scoring approach would improve accuracy by recognising different levels of geographic detail.

Commodity discrepancies occurred because web data often listed only primary commodities, while our database included by-products. For example, Grasberg Mine was recorded as producing copper, gold, silver, and molybdenum, whereas web results mentioned only silver. Categorising commodities into primary and secondary groups through prompt refinement would help resolve this.

\subsection{Total Validation Coverage}
To assess RAV's impact, we compute the total validation coverage, which measures the proportion of assets validated through both LSEG database validation and RAV. Total validation coverage is computed as \( \left( \frac{N_v}{N_t} \right) \times 100 \), where \( N_v \) represents the number of assets validated, and \( N_t \) is the total number of assets in the constructed database.


Table \ref{tab:validation_coverage} presents validation coverage for each company, comparing LSEG-only validation to combined LSEG and RAV validation. Occidental Petroleum (OXY) is excluded due to the absence of company-specific information in the LSEG database and the constructed dataset containing only general assets (e.g., natural gas fields).

Since our validation applies only to named assets, general assets remain largely unverified. While extrapolating validation to unnamed assets could extend coverage, this introduces risks to accuracy and completeness. Notably, RAV significantly increases coverage, underscoring its role in enhancing database robustness by validating assets absent from LSEG.

Coverage varies across companies; D achieves the highest at 33.33\%, while COP has the lowest at 6.43\%, reflecting differences in named asset proportions. Lower coverage suggests a higher proportion of unnamed assets, highlighting gaps in the current validation process.

\begin{table}[H]
    \centering
    \renewcommand{\arraystretch}{1.2} 
    \setlength{\tabcolsep}{3pt} 
    {\fontsize{8}{9}\selectfont 
    \begin{tabular}{|p{1.5cm}|p{1.2cm}|p{2cm}|p{2cm}|} 
        \hline
        \multirow{2}{*}{\textbf{Sector}} & \multirow{2}{*}{\textbf{Company}} & \multicolumn{2}{c|}{\textbf{Validation Coverage}} \\
        \cline{3-4}
        & & \textbf{LSEG Database} & \textbf{LSEG Database + RAV} \\ 
        \hline
        \multirow{5}{*}{Mining} 
        & AA   & 6.96\%  & 20.00\%  \\
        & SCCO & 7.27\%  & 7.27\%   \\
        & FCX  & 7.35\%  & 21.32\%  \\
        & HL   & 17.11\% & 21.05\%  \\
        & NEM  & 19.51\% & 30.89\%  \\
        \hline
        \multirow{4}{*}{Oil and Gas} 
        & CVX  & 7.61\%  & 20.65\%  \\
        & XOM  & 7.69\%  & 23.08\%  \\
        & MPC  & 15.12\% & 25.58\%  \\
        & COP  & 0.71\%  & 6.43\%   \\
        \hline
        \multirow{5}{*}{Utilities} 
        & D    & 6.06\%  & 33.33\%  \\
        & ED   & 1.94\%  & 17.48\%  \\
        & DUK  & 3.90\%  & 23.38\%  \\
        & EXC  & 9.57\%  & 24.35\%  \\
        & NEE  & 0\%     & 10.77\%  \\
        \hline
    \end{tabular}
    }
    \caption{Validation coverage comparison using LSEG databases alone versus LSEG databases with RAV.}
    \label{tab:validation_coverage}
\end{table}



\begin{figure}[H] 
    \centering
    \includegraphics[width=0.5\textwidth]{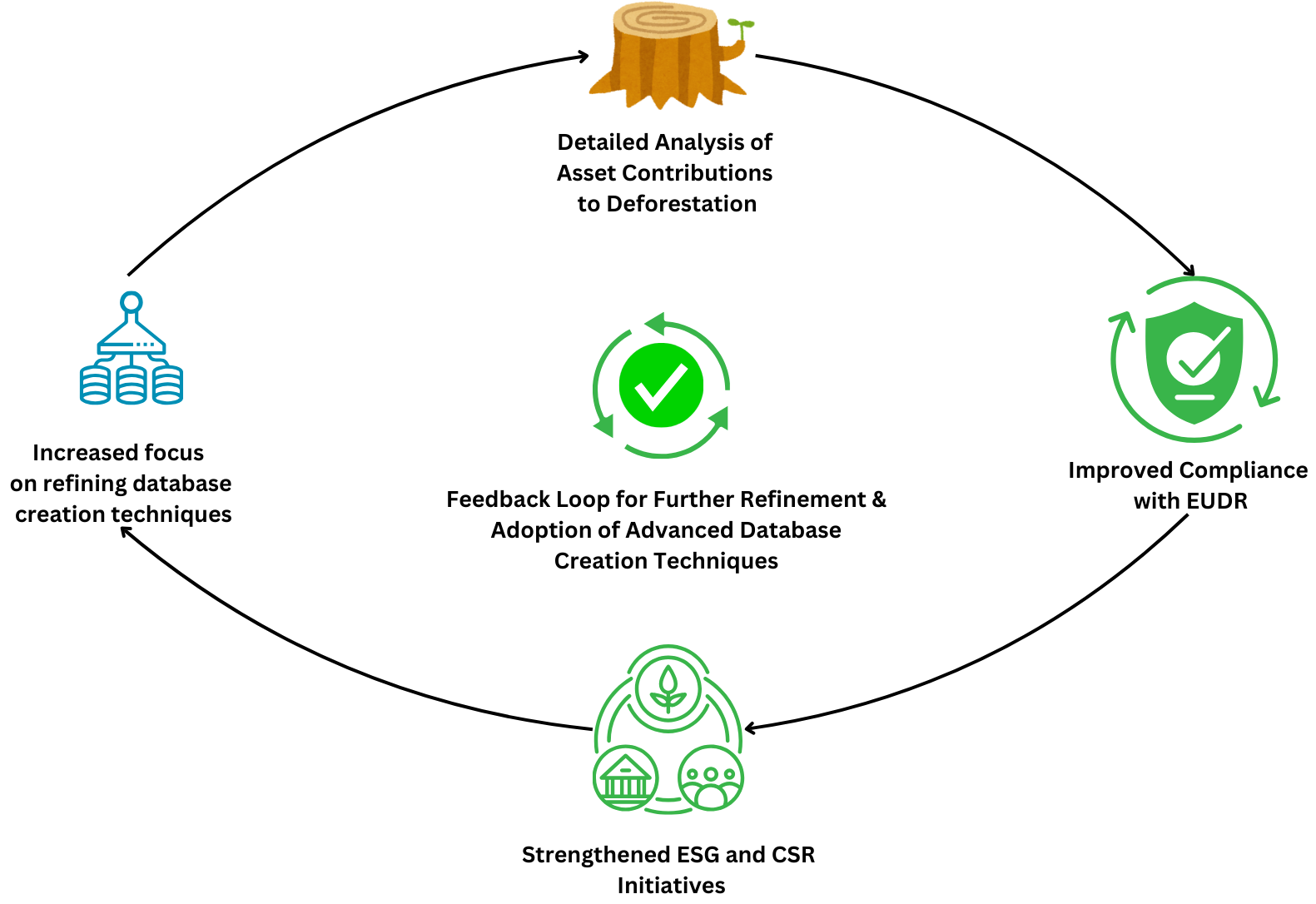} 
    \caption{A feedback loop linking physical asset database creation with improved compliance and ESG
initiatives, driving continuous refinement.}
    \label{fig:feedbackloop}
\end{figure}

As regulatory demands like the EUDR grow, the need for automated, comprehensive databases will increase. Our LLM-based pipeline can adapt to these demands, improving ESG and CSR compliance. The feedback loop (Figure \ref{fig:feedbackloop}) from regulatory success will drive continuous improvements in data quality and database creation techniques, shaping the future of environmental data management.

\section{Conclusion}
This study developed an end-to-end LLM-based pipeline for extracting, structuring, cleaning, and validating physical asset data from SEC EDGAR filings, providing a scalable and automated approach to asset tracking and environmental impact analysis. The pipeline follows a structured architecture, incorporating data acquisition and processing, entity extraction, database creation, database cleaning and a two-step validation framework: LSEG database validation followed by Retrieval-Augmented Validation (RAV). The table parsing improvement module also significantly increased validation coverage, enhancing database completeness. Using LLMs for automation, this study advances structured data extraction and validation, laying the foundation for more efficient regulatory compliance and environmental data management. As regulatory demands evolve, LLM-based techniques will play an increasingly critical role in ensuring accurate, structured asset tracking.

\section{Limitations}
Despite its success, this study has several limitations. The dataset was limited to fifteen companies across three sectors, using only three years of SEC filings, which may restrict the generalisability of the findings. Additionally, the project relied heavily on SEC 10-K filings, excluding potential insights from 8-K, 10-Q, and other reports that may contain relevant asset data. While the IRZ-CoT prompt engineering technique significantly improved entity extraction, its effectiveness in other domains or regulatory environments remains untested. 

The integration of LLMs introduces maintenance and adaptation challenges, especially as newer models may require retuning for continued effectiveness. Biases in pre-trained data could also impact extraction accuracy, particularly in financial and environmental sectors. The RAV validation process depends on web search quality, meaning incomplete or inaccurate online data could affect validation reliability. Although automation reduced manual effort, human verification was necessary for ground truth dataset creation and qualitative assessments. Additionally, the pipeline primarily focuses on historical data, making real-time asset tracking and monitoring challenging.

To address these limitations, future improvements should refine prompt engineering to extract more named assets, reducing reliance on proxies. Expanding the dataset to additional industries and historical filings will enhance coverage and validation effectiveness. Fine-tuning LLMs using domain-specific databases could enhance extraction accuracy. Additionally, dynamic prompt engineering combined with reinforcement learning — where the model is rewarded for accurate extractions and penalized for errors — could help the system adapt more effectively to different types of company disclosures. 

RAV significantly increases validation coverage. It could further benefit from integrating advanced web search APIs and adopting a weighted scoring system to account for partial matches, improving validation granularity. Broader data issues also affected validation when web search results were incomplete or irrelevant, leading to default scores of 0. Expanding search queries, integrating multiple search engines, and refining data extraction techniques could improve RAV’s robustness. Lastly, enhancing automated table parsing would improve structured data extraction from SEC filings, particularly in appendices and financial disclosures where asset details are often tabulated. This could be achieved by automated prompt tailoring, where an LLM (e.g., GPT-4) identifies asset examples from parsed tables and incorporates them into IRZ-CoT prompts.

\section{Ethics Statement}  
This project adheres to the ACM Code of Ethics \cite{acm_code_of_ethics} and the Ethics Guidelines for Trustworthy AI \cite{eu_trustworthy_ai_2019}, emphasizing transparency, fairness, accountability, and technical robustness. It relies on publicly available SEC EDGAR filings, ensuring legal compliance and data integrity. Ethical data acquisition is upheld by adhering to data source terms and conditions, preventing legal conflicts. The project contributes to environmental sustainability by assessing industrial deforestation impacts, supporting global climate initiatives, and reinforcing corporate social responsibility (CSR) through ESG compliance. Accuracy and fairness are prioritized via a two-step validation pipeline, minimizing misrepresentations and addressing LLM biases through transparency and human oversight. By maintaining industry best practices in data processing, the project ensures reliability, robustness, and responsible AI development for ethical and accurate asset tracking.  

 \section*{Acknowledgments}  
I would like to express my sincere gratitude to the LSEG team for providing a free license to LSEG Workspace and for their invaluable guidance throughout the course of this research. Their support significantly contributed to the quality and depth of this work.

\bibliographystyle{acl_natbib}
\bibliography{references}

\begin{thebibliography}{31}
\expandafter\ifx\csname natexlab\endcsname\relax\def\natexlab#1{#1}\fi

\bibitem[{ACM(2018)}]{acm_code_of_ethics}
ACM. 2018.
\newblock \href {https://www.acm.org/code-of-ethics} {Acm code of ethics and professional conduct}.

\bibitem[{Banner et~al.(2019)Banner, Nahshan, Hoffer, and Soudry}]{banner2019posttraining4bitquantizationconvolution}
Ron Banner, Yury Nahshan, Elad Hoffer, and Daniel Soudry. 2019.
\newblock \href {http://arxiv.org/abs/1810.05723} {Post-training 4-bit quantization of convolution networks for rapid-deployment}.

\bibitem[{Beltagy et~al.(2019)Beltagy, Cohan, and Lo}]{scibert}
Iz~Beltagy, Arman Cohan, and Kyle Lo. 2019.
\newblock \href {http://arxiv.org/abs/1903.10676} {Scibert: Pretrained contextualized embeddings for scientific text}.
\newblock \emph{CoRR}, abs/1903.10676.

\bibitem[{Bommarito et~al.(2018)Bommarito, Katz, and Detterman}]{Bommarito2018LexNLP}
M.~Bommarito, D.~Katz, and E.~M. Detterman. 2018.
\newblock \href {https://doi.org/10.2139/ssrn.3192101} {{LexNLP: Natural Language Processing and Information Extraction For Legal and Regulatory Texts}}.
\newblock \emph{InfoSciRN: Legal Informatics (Topic)}.

\bibitem[{Bradley(2020)}]{bradley2020miningforests}
Sian Bradley. 2020.
\newblock \href {https://www.chathamhouse.org/sites/default/files/2020-10/2020-10-14-minings-impacts-forests-bradley.pdf} {Mining's impacts on forests: Aligning policy and finance for climate and biodiversity goals}.
\newblock Technical report, Chatham House, London, UK.

\bibitem[{CGFI(2024)}]{GeoAssetProject2024}
CGFI. 2024.
\newblock \href {https://www.cgfi.ac.uk/spatial-finance-initiative/geoasset-project/petrochemicals/} {Geoasset project: Petrochemicals}.

\bibitem[{Chalkidis et~al.(2020)Chalkidis, Fergadiotis, Malakasiotis, Aletras, and Androutsopoulos}]{legalbert}
Ilias Chalkidis, Manos Fergadiotis, Prodromos Malakasiotis, Nikolaos Aletras, and Ion Androutsopoulos. 2020.
\newblock \href {http://arxiv.org/abs/2010.02559} {{LEGAL-BERT:} the muppets straight out of law school}.
\newblock \emph{CoRR}, abs/2010.02559.

\bibitem[{Chung et~al.(2022)Chung, Hou, Longpre, Zoph, Tay, Fedus, Li, Wang, Dehghani, Brahma, Webson, Gu, Dai, Suzgun, Chen, Chowdhery, Castro-Ros, Pellat, Robinson, Valter, Narang, Mishra, Yu, Zhao, Huang, Dai, Yu, Petrov, Chi, Dean, Devlin, Roberts, Zhou, Le, and Wei}]{chung2022scalinginstructionfinetunedlanguagemodels}
Hyung~Won Chung, Le~Hou, Shayne Longpre, Barret Zoph, Yi~Tay, William Fedus, Yunxuan Li, Xuezhi Wang, Mostafa Dehghani, Siddhartha Brahma, Albert Webson, Shixiang~Shane Gu, Zhuyun Dai, Mirac Suzgun, Xinyun Chen, Aakanksha Chowdhery, Alex Castro-Ros, Marie Pellat, Kevin Robinson, Dasha Valter, Sharan Narang, Gaurav Mishra, Adams Yu, Vincent Zhao, Yanping Huang, Andrew Dai, Hongkun Yu, Slav Petrov, Ed~H. Chi, Jeff Dean, Jacob Devlin, Adam Roberts, Denny Zhou, Quoc~V. Le, and Jason Wei. 2022.
\newblock \href {http://arxiv.org/abs/2210.11416} {Scaling instruction-finetuned language models}.

\bibitem[{Dettmers et~al.(2023)Dettmers, Pagnoni, Holtzman, and Zettlemoyer}]{dettmers2023qloraefficientfinetuningquantized}
Tim Dettmers, Artidoro Pagnoni, Ari Holtzman, and Luke Zettlemoyer. 2023.
\newblock \href {http://arxiv.org/abs/2305.14314} {Qlora: Efficient finetuning of quantized llms}.

\bibitem[{Deußer et~al.(2022)Deußer, Ali, Hillebrand, Nurchalifah, Jacob, Bauckhage, and Sifa}]{Deuer2022}
Tobias Deußer, Syed~Musharraf Ali, Lars Hillebrand, Desiana Nurchalifah, Basil Jacob, Christian Bauckhage, and Rafet Sifa. 2022.
\newblock \href {https://doi.org/10.1109/icmla55696.2022.00254} {Kpi-edgar: A novel dataset and accompanying metric for relation extraction from financial documents}.
\newblock In \emph{2022 21st IEEE International Conference on Machine Learning and Applications (ICMLA)}. IEEE.

\bibitem[{Devlin et~al.(2018)Devlin, Chang, Lee, and Toutanova}]{DBLP:journals/corr/abs-1810-04805}
Jacob Devlin, Ming{-}Wei Chang, Kenton Lee, and Kristina Toutanova. 2018.
\newblock \href {http://arxiv.org/abs/1810.04805} {{BERT:} pre-training of deep bidirectional transformers for language understanding}.
\newblock \emph{CoRR}, abs/1810.04805.

\bibitem[{ETC(2024)}]{EnergyTransitions}
ETC. 2024.
\newblock \href {https://www.energy-transitions.org/financing-the-transition-the-costs-of-avoiding-deforestation/} {Financing the transition: The costs of avoiding deforestation}.
\newblock Technical report, Energy~Transitions~Commission, London, UK.

\bibitem[{{European Commission}(2019)}]{eu_trustworthy_ai_2019}
{European Commission}. 2019.
\newblock \href {https://digital-strategy.ec.europa.eu/en/library/ethics-guidelines-trustworthy-ai} {Ethics guidelines for trustworthy ai}.

\bibitem[{{European Commission}(2023)}]{EUDR}
{European Commission}. 2023.
\newblock \href {https://environment.ec.europa.eu/topics/forests/deforestation/regulation-deforestation-free-products_en} {Regulation on deforestation-free products}.

\bibitem[{Finer et~al.(2008)Finer, Jenkins, Pimm, Keane, and Ross}]{Finer2008}
Matt Finer, Clinton~N. Jenkins, Stuart~L. Pimm, Brian Keane, and Carl Ross. 2008.
\newblock \href {https://doi.org/10.1371/journal.pone.0002932} {Oil and gas projects in the western amazon: Threats to wilderness, biodiversity, and indigenous peoples}.
\newblock \emph{PLoS ONE}, 3(8):e2932.

\bibitem[{{Forest 500}(2024)}]{forest5002024}
{Forest 500}. 2024.
\newblock \href {https://forest500.org/publications/2024-a-decade-of-deforestation-data/} {2024: A decade of deforestation data}.

\bibitem[{{Google Developers}(2024)}]{google_custom_search_api}
{Google Developers}. 2024.
\newblock \href {https://developers.google.com/custom-search/v1/overview} {Custom search json api overview}.

\bibitem[{Hu et~al.(2024)Hu, Yu, Chen, and Ponti}]{hu2024finetuninglargelanguagemodels}
Hanxu Hu, Simon Yu, Pinzhen Chen, and Edoardo~M. Ponti. 2024.
\newblock \href {http://arxiv.org/abs/2403.07794} {Fine-tuning large language models with sequential instructions}.

\bibitem[{Imperiale et~al.(2023)Imperiale, Pizzi, and Lippolis}]{IMPERIALE2023101468}
Francesca Imperiale, Simone Pizzi, and Stella Lippolis. 2023.
\newblock \href {https://doi.org/https://doi.org/10.1016/j.jup.2022.101468} {Sustainability reporting and esg performance in the utilities sector}.
\newblock \emph{Utilities Policy}, 80:101468.

\bibitem[{IntegrityNext(2024)}]{integritynext2024energyutilities}
IntegrityNext. 2024.
\newblock \href {https://www.integritynext.com/industries/energy-utilities} {Energy {\&} utilities}.

\bibitem[{Jiang et~al.(2023)Jiang, Xu, Gao, Sun, Liu, Dwivedi-Yu, Yang, Callan, and Neubig}]{jiang2023active}
Zhengbao Jiang, Frank~F. Xu, Luyu Gao, Zhiqing Sun, Qian Liu, Jane Dwivedi-Yu, Yiming Yang, Jamie Callan, and Graham Neubig. 2023.
\newblock \href {http://arxiv.org/abs/2305.06983} {Active retrieval augmented generation}.

\bibitem[{Lee et~al.(2019)Lee, Yoon, Kim, Kim, Kim, So, and Kang}]{biobert}
Jinhyuk Lee, Wonjin Yoon, Sungdong Kim, Donghyeon Kim, Sunkyu Kim, Chan~Ho So, and Jaewoo Kang. 2019.
\newblock \href {https://doi.org/10.1093/bioinformatics/btz682} {{BioBERT: a pre-trained biomedical language representation model for biomedical text mining}}.
\newblock \emph{Bioinformatics}, 36(4):1234--1240.

\bibitem[{{London Stock Exchange Group (LSEG)}(2024)}]{LSEG_Workspace}
{London Stock Exchange Group (LSEG)}. 2024.
\newblock \href {https://www.lseg.com/en/data-analytics/products/workspace} {Lseg workspace: Data and analytics}.

\bibitem[{{Moody et al.}(2024)}]{sec-edgar}
{Moody et al.} 2024.
\newblock \href {https://github.com/sec-edgar/sec-edgar} {Sec-edgar: Download all companies' periodic reports, filings, and forms from edgar database}.
\newblock GitHub repository, Apache-2.0 License.

\bibitem[{Reimers and Gurevych(2019)}]{reimers-2019-sentence-bert}
Nils Reimers and Iryna Gurevych. 2019.
\newblock \href {http://arxiv.org/abs/1908.10084} {Sentence-bert: Sentence embeddings using siamese bert-networks}.
\newblock In \emph{Proceedings of the 2019 Conference on Empirical Methods in Natural Language Processing}. Association for Computational Linguistics.

\bibitem[{Rosenberg et~al.(2000)Rosenberg, McCully, and Pringle}]{pringle}
David~M. Rosenberg, Patrick McCully, and Catherine~M. Pringle. 2000.
\newblock \href {https://doi.org/10.1641/0006-3568(2000)050[0746:GSEEOH]2.0.CO;2} {Global-scale environmental effects of hydrological alterations: Introduction}.
\newblock \emph{BioScience}, 50(9):746--751.

\bibitem[{Tjong Kim~Sang and De~Meulder(2003)}]{tjong-kim-sang-de-meulder-2003-introduction}
Erik~F. Tjong Kim~Sang and Fien De~Meulder. 2003.
\newblock \href {https://www.aclweb.org/anthology/W03-0419} {Introduction to the {C}o{NLL}-2003 shared task: Language-independent named entity recognition}.
\newblock In \emph{Proceedings of the Seventh Conference on Natural Language Learning at {HLT}-{NAACL} 2003}, pages 142--147.

\bibitem[{Trajanoska et~al.(2023)Trajanoska, Stojanov, and Trajanov}]{trajanoska2023enhancingknowledgegraphconstruction}
Milena Trajanoska, Riste Stojanov, and Dimitar Trajanov. 2023.
\newblock \href {http://arxiv.org/abs/2305.04676} {Enhancing knowledge graph construction using large language models}.

\bibitem[{Wang et~al.(2023)Wang, Sun, Li, Ouyang, Wu, Zhang, Li, and Wang}]{wang2023gptnernamedentityrecognition}
Shuhe Wang, Xiaofei Sun, Xiaoya Li, Rongbin Ouyang, Fei Wu, Tianwei Zhang, Jiwei Li, and Guoyin Wang. 2023.
\newblock \href {http://arxiv.org/abs/2304.10428} {Gpt-ner: Named entity recognition via large language models}.

\bibitem[{Watch(2016)}]{amazonwatch2016crude}
Amazon Watch. 2016.
\newblock \href {https://amazonwatch.org/assets/files/2016-amazon-crude-report.pdf} {The amazon crude: How american consumers fuel deforestation and human rights abuses in south america}.
\newblock Technical report, Amazon Watch, Oakland, CA.

\bibitem[{Zhao et~al.(2023)Zhao, Yilahun, and Hamdulla}]{pipelinecot}
Hangtian Zhao, Hakiz Yilahun, and Askar Hamdulla. 2023.
\newblock \href {https://doi.org/10.1109/IALP61005.2023.10337264} {Pipeline chain-of-thought: A prompt method for large language model relation extraction}.
\newblock In \emph{2023 International Conference on Asian Language Processing (IALP)}, pages 31--36.

\end{thebibliography}


\newpage
\appendix

\section{Appendix}
\label{sec:appendix}

\subsection{Sectors, Companies and Stock Tickers used}
\label{stocktickers}

\begin{table}[H]
    \centering
    \resizebox{\columnwidth}{!}{%
    \renewcommand{\arraystretch}{1.2} 
    \begin{tabular}{|l|l|c|}
        \hline
        \textbf{Sector} & \textbf{Company Name} & \textbf{Stock Ticker} \\ 
        \hline
        \multirow{5}{*}{Mining} 
        & Alcoa Corporation & AA \\ 
        & Hecla Mining Corporation & HL \\ 
        & Newmont Corporation & NEM \\ 
        & Freeport Mc-Moran & FCX \\ 
        & Southern Copper Corporation & SCCO \\ 
        \hline
        \multirow{5}{*}{Oil \& Gas} 
        & ConocoPhillips Company & COP \\ 
        & Marathon Petroleum Corporation & MPC \\ 
        & Chevron Corporation & CVX \\ 
        & Occidental Petroleum & OXY \\ 
        & Exxon Mobil Corporation & XOM \\ 
        \hline
        \multirow{5}{*}{Utilities} 
        & Dominion Energy & D \\ 
        & Duke Energy Corporation & DUK \\ 
        & Consolidated Edison & ED \\ 
        & Exelon Corporation & EXC \\ 
        & NextEra Energy & NEE \\ 
        \hline
    \end{tabular}%
    }
    \caption{Companies used for pipeline construction, grouped by sector.}
    \label{tab:companies}
\end{table}

\subsection{Chunks per document in Chunk-based Querying Technique}
\begin{figure}[H]
    \centering
    \includegraphics[width=1.0\linewidth]{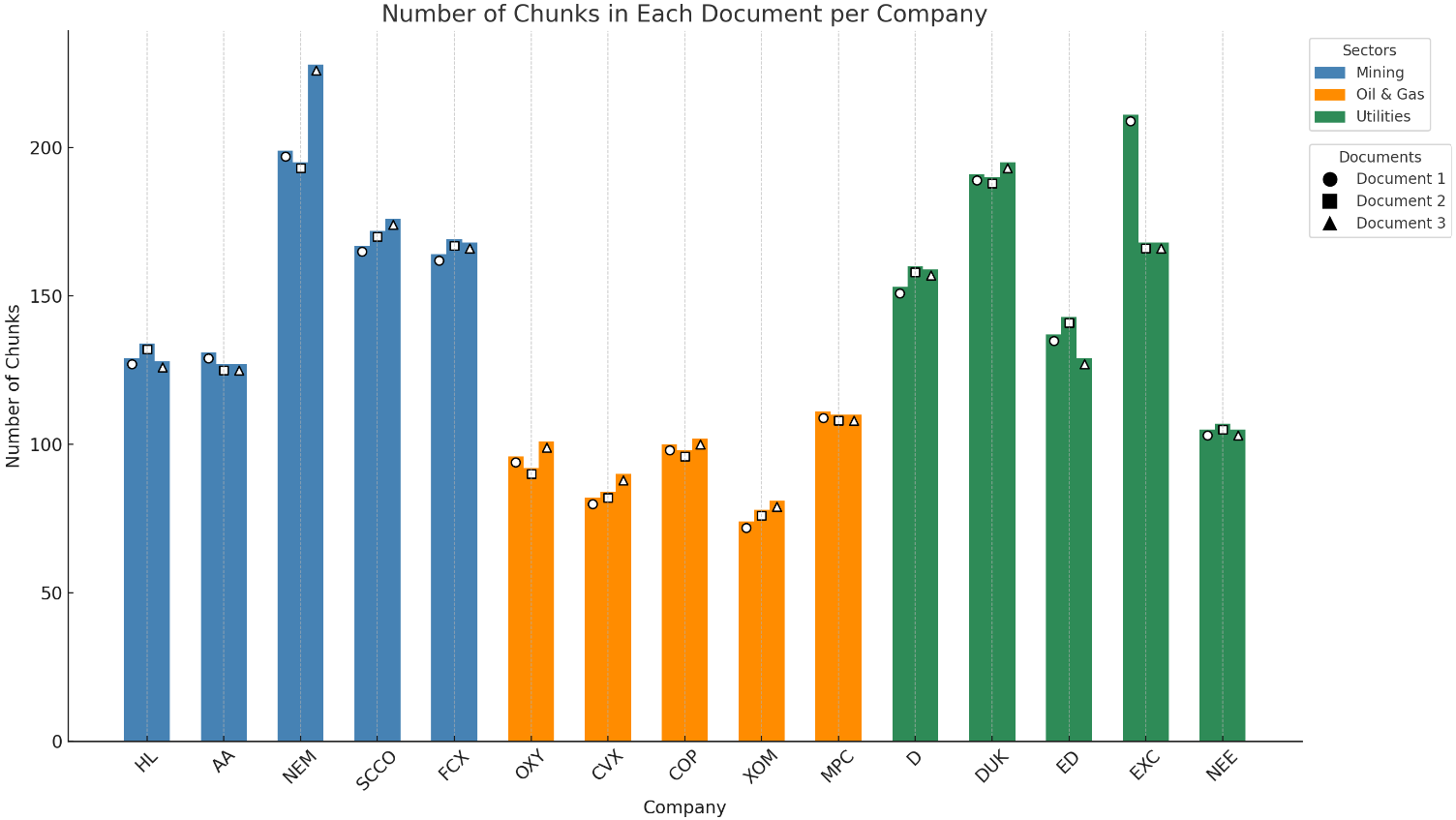}
    \caption{Number of chunks generated per document for each company across the three sectors.}
    \label{fig:chunkspercomp}
\end{figure}

\subsection{Cosine similarities of LLM-generated outputs against NER results}
\label{cosinesimilaritiesLLMNER}

\begin{table}[H]
    \centering
    \resizebox{\columnwidth}{!}{%
    \small
    \renewcommand{\arraystretch}{1.2}
    \begin{tabular}{|l|c|c|}
        \hline
        \textbf{Model} & \textbf{Location - cosine similarity} & \textbf{Organisation - cosine similarity} \\
        \hline
        Mistral-7B & 0.6945 & 0.6809 \\
        Llama 3    & 0.5987 & 0.7177 \\
        Gemma 2    & 0.7702 & 0.7461 \\
        \hline
    \end{tabular}
    }
    \caption{Cosine similarities of LLM-generated outputs against NER results.}
    \label{tab:llmvsner}
\end{table}

\subsection{Error Analysis of LLM-generated outputs against NER results}
\label{erroranalysisLLMNER}
\begin{table}[H]
    \centering
    \small
    \renewcommand{\arraystretch}{1.2}
    \begin{tabular}{|p{0.16\columnwidth}|p{0.16\columnwidth}|p{0.60\columnwidth}|}
        \hline
        \textbf{Model} & \textbf{Type} & \textbf{Description (LLM vs. NER Output)} \\ 
        \hline
        Mistral-7B & Success & \textbf{LLM:} Accurately extracted locations “South Church Street, Charlotte, North Carolina” and organisations. \newline
        \textbf{NER:} Fragmented these entities into “Securities Exchange” and “\#\#TON.” Split up locations like ‘US’ into ‘U’ and ‘S’. \\ \hline
                   & Failure & \textbf{LLM:} Sometimes failed to extract any entities in the text. \newline
        \textbf{NER:} Correctly identified “Duke Energy.” \\ \hline
                   & Challenge & \textbf{LLM:} Provided a list of relevant organisations but failed to identify important locations, leading to lower location scores. \newline
        \textbf{NER:} Captured more entities but included irrelevant fragments like “GENERA” and “Deer Holding Corp.” \\ \hline
        LLaMA 3   & Success & \textbf{LLM:} Successfully identified the organisation “Duke Energy” without errors or fragmentation. \newline
        \textbf{NER:} Also correctly extracted “Duke Energy” but severely fragmented other entities by outputting locations like “U.Securities” and “. S”. Provided incorrect outputs including “Per” and “Board of Directors”. \\ \hline
                   & Failure & \textbf{LLM:} Sometimes failed to extract any locations or organisations, resulting in no output. \newline
        \textbf{NER:} Identified several locations like “Central” and “Gulf Coast” and organisations such as “Spectra Energy” and “ConocoPhillips.” \\ \hline
                   & Challenge & \textbf{LLM:} Identified relevant organisations and locations, but the output was marred by repeated and irrelevant text. \newline
        \textbf{NER:} Also extracted similar entities but with many fragmented outputs, lowering accuracy. \\ \hline
        Gemma 2  & Success & \textbf{LLM:} Successfully identified “South Church Street, Charlotte, North Carolina” and several organisations. \newline
        \textbf{NER:} Produced fragmented entities like “Securities Exchange” and “\#\#TON,” and returned locations like “\#\#mian”, “South” and “Central”. \\ \hline
                   & Failure & \textbf{LLM:} Sometimes did not extract any entities. \newline
        \textbf{NER:} Identified locations like “Midwestern United States” and organisations such as “Duke Energy” and “Cinergy.” \\ \hline
                   & Challenge & \textbf{LLM:} Sometimes did not extract any locations. \newline
        \textbf{NER:} Captured a broader range of organisations but included several irrelevant fragments. \\ \hline
    \end{tabular}
    \caption{Examples of Successes, Failures, and Challenges in entity extraction by LLMs compared to NER.}
    \label{tab:llm_analysis}
\end{table}

\clearpage
\newpage
\subsection{Problem descriptions in Iterative Prompt Refinement}
\label{promptproblem}
\begin{table}[H]
    \centering
    \resizebox{\columnwidth}{!}{%
    \small
    \renewcommand{\arraystretch}{1.2}
    \begin{tabular}{|p{0.3\columnwidth}|p{0.35\columnwidth}|p{0.35\columnwidth}|}
        \hline
        \rowcolor{lightgray}\textbf{Prompt Name} & \textbf{Problem Description} & \textbf{Problem Example} \\ \hline
        Zero-shot & Incorrect mention of financial assets as physical assets, lack of location specificity & \texttt{`asset: Consolidated Operations - Revenues', `location: Seven countries'} \\ \hline
        \rowcolor{lightgray}One-shot & Hallucination --- includes example in answer & \texttt{`ownership: Freeport-McMoran'} \\ \hline
        Few-shot & Incorrect classification of physical asset, hallucination --- includes example in answer & \texttt{`physical asset: Granted patents (intellectual property), Registered trademarks (intellectual property)', `physical assets: [Grasberg mine]'} \\ \hline
        \rowcolor{lightgray}Chain-of-thought & Incorrect mention of location as physical asset; provides irrelevant answers if no information is available & \texttt{`physical asset: Brazil', `relationships: [asset: '', location: '', ownership: '', commodities: '']'} \\ \hline
        Generated knowledge & Outputs overly detailed and verbose answers & \texttt{`physical assets: [substantially all assets of the Company, ...]'} \\ \hline
        \rowcolor{lightgray}Prompt chaining & Lack of specificity & \texttt{`physical assets: [substantially all assets of the Company]', `physical assets: [mining operations, properties, leases]'} \\ \hline
        Role prompting & Improvement noted but lack of specificity & \texttt{`physical assets: [Mining operations, processing plants, manufacturing facilities]'} \\ \hline
        \rowcolor{lightgray}Role and instructional prompting & Improvement noted but lack of specificity & \texttt{`physical assets: [facilities]'} \\ \hline
    \end{tabular}
    }
    \caption{Problem descriptions and examples for each prompting technique.}
    \label{tab:prompt_techniques}
\end{table}

\subsection{Computational times of prompting techniques in Iterative Prompt Refinement}
\label{prompttime}
\begin{figure}[H]
  \centering
  \includegraphics[width=1.0\linewidth]{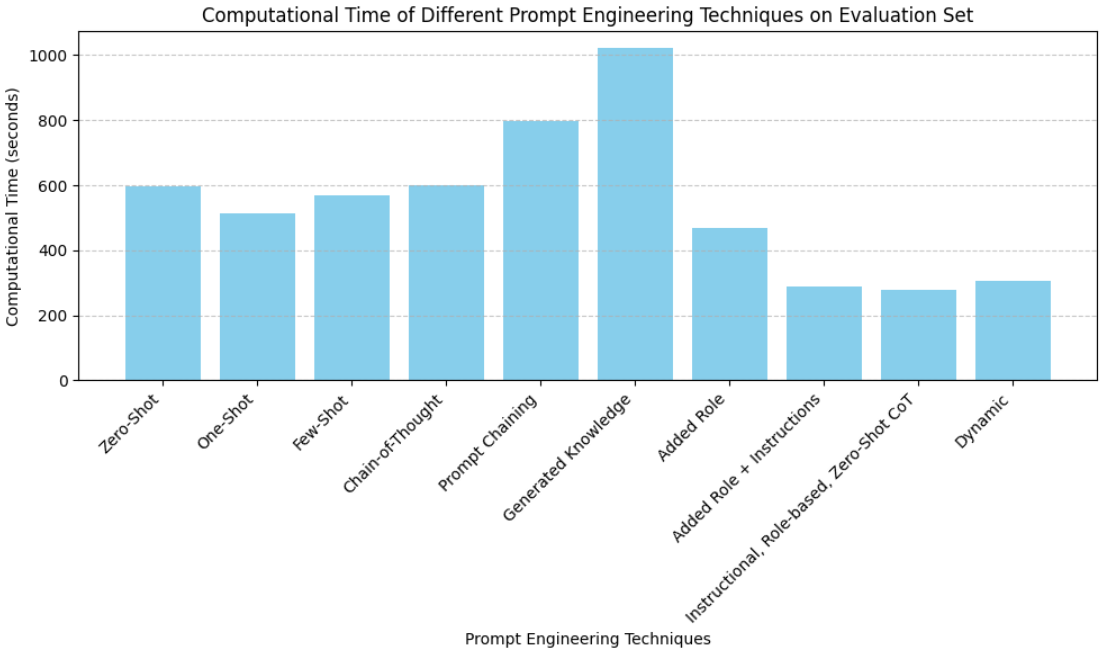}
  \caption{Computational time (in seconds) required for different prompt engineering techniques on the evaluation set.}
  \label{fig:prompttime}
\end{figure}

\subsection{Results of LLM Ensemble Implementation}
\label{ensembleresults}
\begin{figure}[H]
  \centering
  \includegraphics[width=1.0\linewidth]{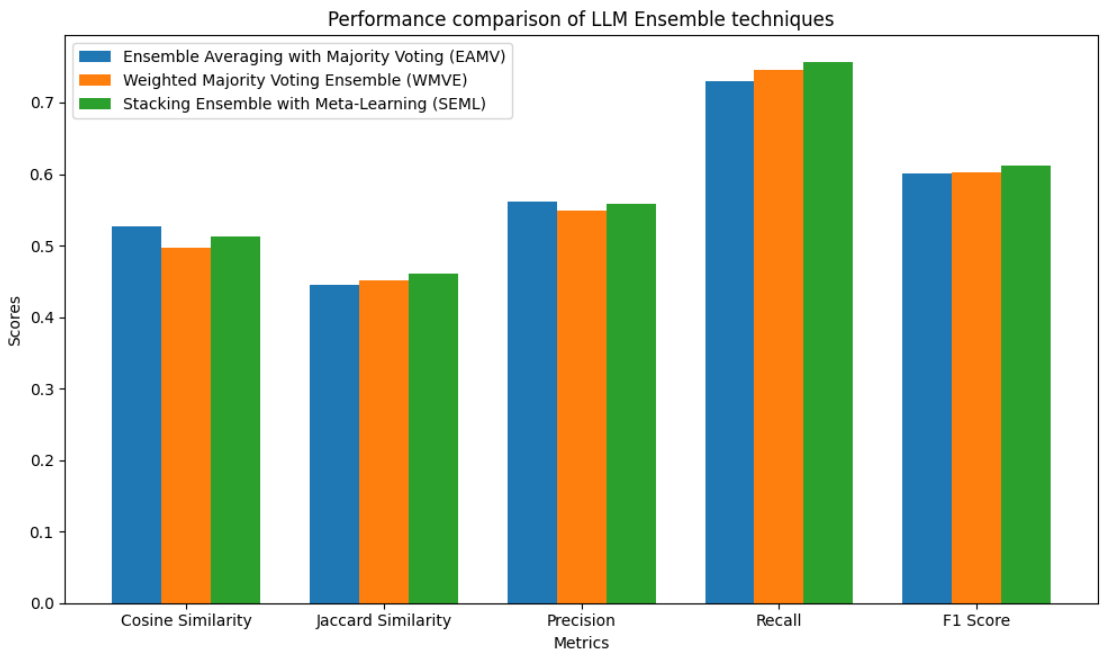} 
  \caption{Performance comparison of various LLM ensemble techniques.}
  \label{fig:ensembleresults}
\end{figure}

\subsection{Efficiency of LLM Ensemble Methods}
\label{ensembletimes}
\vspace*{-\baselineskip}
\begin{figure}[H]
\captionsetup{skip=0pt}
  \centering
  \includegraphics[width=1.0\linewidth]{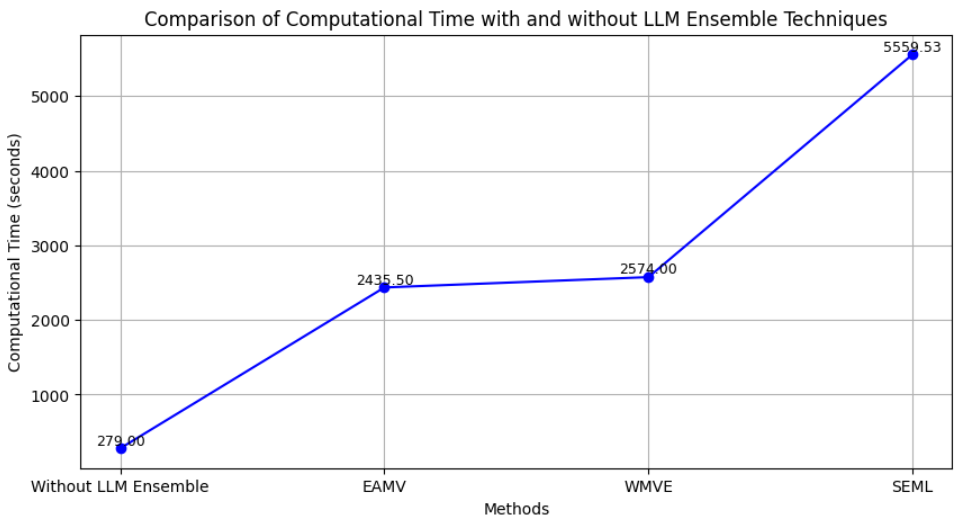} 
  \caption{Comparison of computational times with and without using LLM ensemble techniques.}
  \label{fig:ensembletimes}
\end{figure}

\subsection{LSEG Database Validation Results}
\label{LSEGvalresults}
\begin{figure}[H]
    \centering
    \includegraphics[width=0.9\linewidth]{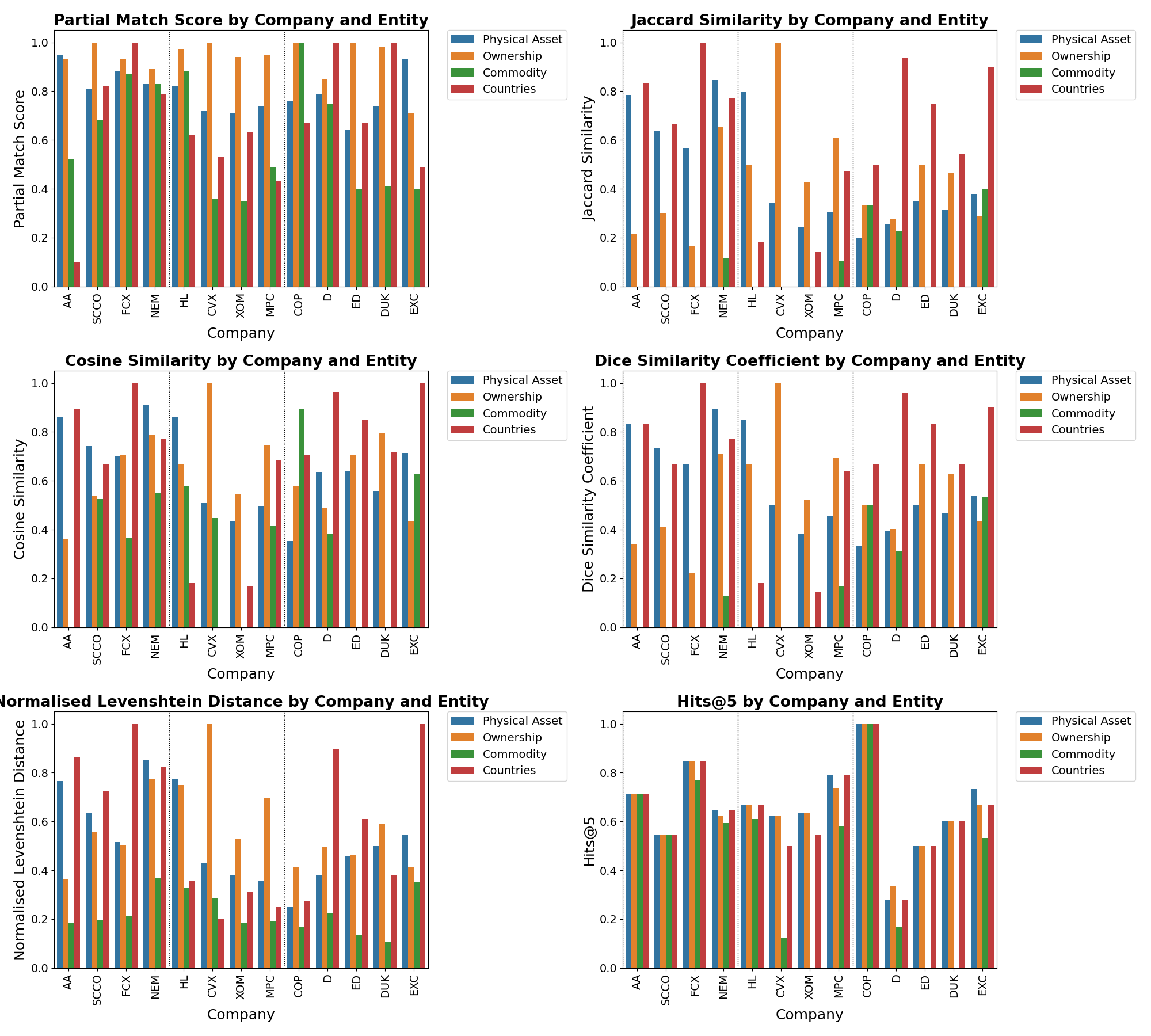}
    \caption{Results from the LSEG database validation across various metrics.}
    \label{fig:LSEGvalresults}
\end{figure}

\subsection{Coverage Calculation Results}
\label{coverageresults}
\begin{table}[H]
    \centering
    \small
    \renewcommand{\arraystretch}{1.2}
    \begin{tabular}{|c|c|c|}
        \hline
        \textbf{Sector} & \textbf{Company} & \textbf{Coverage} \\ \hline
        \multirow{5}{*}{Mining} 
            & AA   & 71.43\% \\ \cline{2-3}
            & SCCO & 54.55\% \\ \cline{2-3}
            & HL   & 66.67\% \\ \cline{2-3}
            & NEM  & 62.16\% \\ \cline{2-3}
            & FCX  & 92.31\% \\ \hline
        \multirow{5}{*}{Oil \& Gas} 
            & CVX & 25\% \\ \cline{2-3}
            & XOM & 0\% \\ \cline{2-3}
            & MPC & 94.74\% \\ \cline{2-3}
            & OXY & N/A \\ \cline{2-3}
            & COP & 100\% \\ \hline
        \multirow{5}{*}{Utilities} 
            & D   & 27.78\% \\ \cline{2-3}
            & DUK & 40\% \\ \cline{2-3}
            & EXC & 13.33\% \\ \cline{2-3}
            & ED  & 50\% \\ \cline{2-3}
            & NEE & N/A \\ \hline
    \end{tabular}
    \caption{Calculated coverage percentages for databases across sectors.}
    \label{tab:coverage_table}
\end{table}

\subsection{Coverage before and after implementation of table parsing}
\label{coveragetableparse}
\begin{figure}[H]
  \centering
  \includegraphics[width=1.0\linewidth]{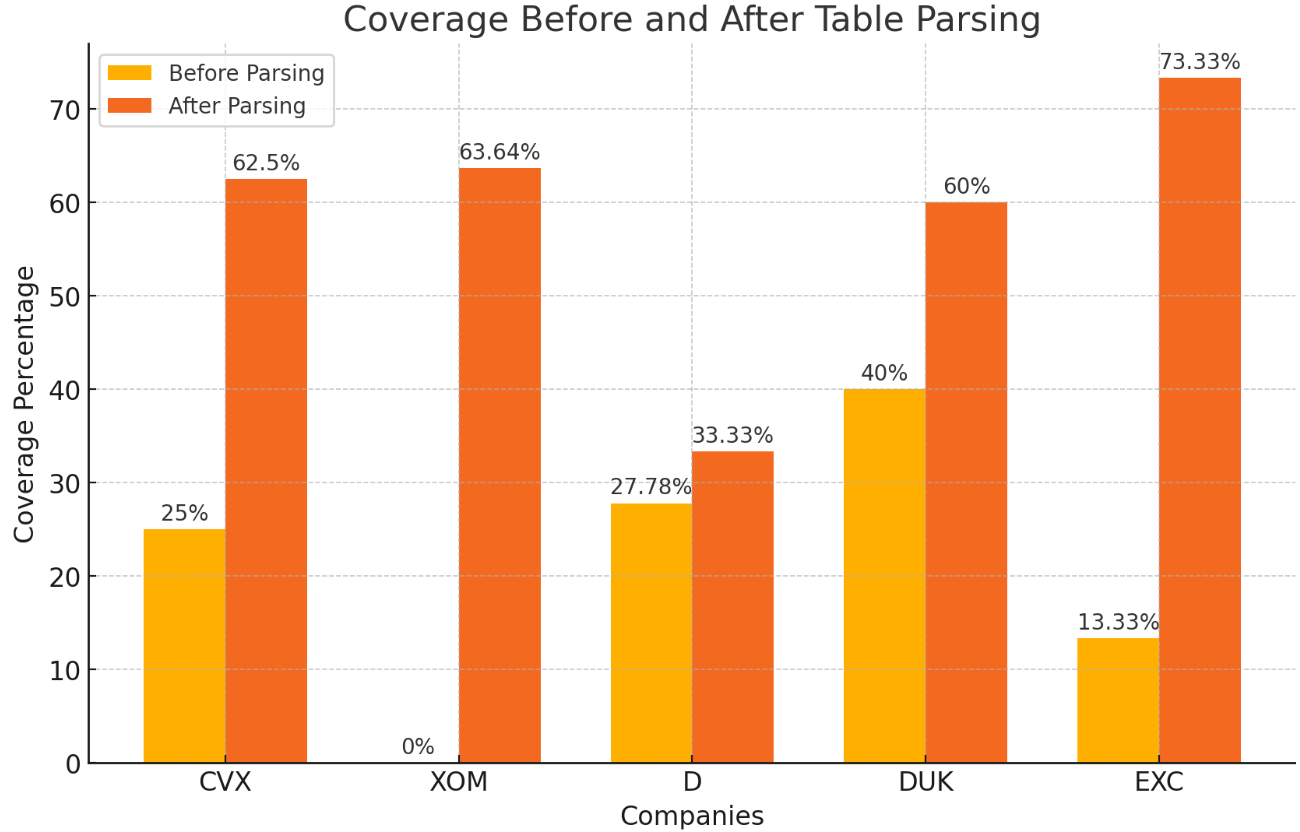}
  \caption{Coverage percentages before and after performing table parsing.}
  \label{fig:coveragebeforeaftertableparse}
\end{figure}

\subsection{Retrieval Augmented Validation (RAV) Results}
\label{RAVresults}
\begin{table}[H]
    \centering
    \small
    \renewcommand{\arraystretch}{1.2}
    \begin{tabular}{|l|l|c|}
        \hline
        \textbf{Sector} & \textbf{Company} & \textbf{Similarity Score} \\ \hline
        \multirow{5}{*}{Mining} 
            & AA   & 0.42 \\ \cline{2-3}
            & SCCO & 0.37 \\ \cline{2-3}
            & FCX  & 0.43 \\ \cline{2-3}
            & NEM  & 0.38 \\ \cline{2-3}
            & HL   & 0.46 \\ \hline
        \multirow{4}{*}{Oil and Gas} 
            & CVX  & 0.44 \\ \cline{2-3}
            & XOM  & 0.48 \\ \cline{2-3}
            & MPC  & 0.52 \\ \cline{2-3}
            & COP  & 0.52 \\ \hline
        \multirow{5}{*}{Utilities} 
            & D    & 0.33 \\ \cline{2-3}
            & ED   & 0.38 \\ \cline{2-3}
            & DUK  & 0.40 \\ \cline{2-3}
            & EXC  & 0.31 \\ \cline{2-3}
            & NEE  & 0.57 \\ \hline
    \end{tabular}
    \caption{Averaged classification similarity scores from Retrieval Augmented Validation (RAV).}
    \label{tab:RAVresults}
\end{table}


\subsection{Prompt Library}
\label{sec:promptlibrary}

In this section, we present the comprehensive collection of prompts utilised throughout this project for information extraction. The prompt library consists of a variety of carefully designed instructions aimed at guiding LLMs in extracting specific entities and relationships, such as physical assets, locations, ownership details, and commodities, from SEC EDGAR filings. Each prompt is tailored to enhance the performance of LLMs in different scenarios, employing techniques such as zero-shot, one-shot, few-shot prompting, and more sophisticated methods like Chain-of-Thought (CoT) reasoning and role-based prompting, before constructing our IRZ-CoT prompt. \newline

The prompts are categorised based on their usage in different stages of the project, including entity extraction, database cleaning and when creating an improvement module. This prompt library serves as the foundation for automating the extraction process and ensuring the reliability and accuracy of the data. Each prompt has been optimised through iterative testing and refinement to address the unique challenges posed by our use case.

\begin{tcolorbox}[breakable, colback=white!5, colframe=black!75, title=Prompt for extracting locations and organisations when comparing LLMs and NER]
\begin{lstlisting}[basicstyle=\small\ttfamily, breaklines=true]
    prompt_instruction = "You are a virtual assistant with advanced expertise in a broad spectrum of topics, equipped to utilize high-level critical thinking, cognitive skills, creativity, and innovation. Your goal is to deliver the most straightforward and accurate answer possible for each question, ensuring high-quality and useful responses for the user."
    user_prompt = f"Text: {chunk}\nQuery: Does this text mention any locations or organisations? If yes, please specify them in the following format:\nlocations: [ ]\norganisations: [ ]"
\end{lstlisting}
\end{tcolorbox}

\begin{tcolorbox}[breakable, colback=white!5, colframe=black!75, title=Zero-shot]
\begin{lstlisting}[basicstyle=\small\ttfamily, breaklines=true]
f"Text: {chunk}\nQuery: Does this text mention any physical assets, locations, ownerships, and commodities? "
        "If yes, please specify them in the following format:\n"
        "physical assets: [ ]\nlocations: [ ]\nownerships: [ ]\ncommodities: []\n"
        "Additionally, identify the relationships between them, specifying the location of each physical asset, their ownership details, and commodities. "
        "Format the relationships as follows:\nrelationships: [asset: '', location: '', ownership: '', commodities: '']"
\end{lstlisting}
\end{tcolorbox}

\begin{tcolorbox}[breakable, colback=white!5, colframe=black!75, title=One-shot]
\begin{lstlisting}[basicstyle=\small\ttfamily, breaklines=true]
 f"Text: {chunk}\nQuery: Does this text mention any physical assets, locations, ownerships, and commodities? "
        "If yes, please specify them in the following format:\n"
        "physical assets: [ ]\nlocations: [ ]\nownerships: [ ]\ncommodities: []\n"
        "Additionally, identify the relationships between them, specifying the location of each physical asset, their ownership details, and commodities. "
        "Format the relationships as follows:\nrelationships: [asset: '', location: '', ownership: '', commodities: '']"
        "Here is an example:\n"
        "Example:\n"
        "Text: [...] Our principal asset is the Grasberg mine, which we discovered in 1988. Grasberg contains the largest single gold reserve and one of the largest copper reserves of any mine in the world. Our principal operating subsidiary is PT Freeport Indonesia, a limited liability company organized under the laws of the Republic of Indonesia and incorporated in Delaware. [...]"
        "Query: Does this text mention any physical assets, locations, and ownerships?\n"
        "physical assets: [Grasberg mine]\nlocations: [Sudirman Mountain Range, Papua, Indonesia]\nownerships: [Republic of Indonesia, Delaware]\n[commodities: copper, gold]\n"
        "relationships: [asset: 'Grasberg mine', location: 'Indonesia', ownership: 'PT Freeport Indonesia', commodities: 'copper', 'gold']\n\n"
\end{lstlisting}
\end{tcolorbox}

\begin{tcolorbox}[breakable, colback=white!5, colframe=black!75, title=Few-shot]
\begin{lstlisting}[basicstyle=\small\ttfamily, breaklines=true]
        f"Text: {chunk}\nQuery: Does this text mention any physical assets, locations, ownerships, and commodities? "
        "If yes, please specify them in the following format:\n"
        "physical assets: [ ]\nlocations: [ ]\nownerships: [ ]\ncommodities: []\n"
        "Additionally, identify the relationships between them, specifying the location of each physical asset, their ownership details, and commodities. "
        "Format the relationships as follows:\nrelationships: [asset: '', location: '', ownership: '', commodities: '']"
        "Here are some examples:\n"
        "Example:\n"
        "Text: [...] Our principal asset is the Grasberg mine, which we discovered in 1988. Grasberg contains the largest single gold reserve and one of the largest copper reserves of any mine in the world. Our principal operating subsidiary is PT Freeport Indonesia, a limited liability company organized under the laws of the Republic of Indonesia and incorporated in Delaware. [...]"
        "Query: Does this text mention any physical assets, locations, and ownerships?\n"
        "physical assets: [Grasberg mine]\nlocations: [Sudirman Mountain Range, Papua, Indonesia]\nownerships: [Republic of Indonesia, Delaware]\n[commodities: copper, gold]\n"
        "relationships: [asset: 'Grasberg mine', location: 'Indonesia', ownership: 'PT Freeport Indonesia', commodities: 'copper', 'gold']\n\n"

        "Example 2:\n"
        "Text: [...] PT Freeport Indonesia mines, processes and explores for ore containing copper, gold and silver. It operates in the remote highlands of the Sudirman Mountain Range in the province of Papua (formerly Irian Jaya), Indonesia, which is on the western half of the island of New Guinea. [...]"
        "Query: Does this text mention any physical assets, locations, and ownerships?\n"
        "physical assets: [PT Freeport Indonesia Mines]\nlocations: [Sudirman Mountain Range, Papua, Indonesia, New Guinea]\nownerships: [PT Freeport Indonesia]\ncommodities: [copper, gold, silver]\n"
        "relationships: \n"
        "[asset: 'PT Freeport Indonesia Mines', location: 'Sudirman Mountain Range, Papua, Indonesia, New Guinea', ownership: 'PT Freeport Indonesia', commodities: 'copper, gold, silver']\n"

        "Example 3:\n"
        "Text: [...] The Republic of Indonesia consists of more than 17,000 islands stretching 3,000 miles along the equator from Malaysia to Australia and is the fourth most populous nation in the world with over 200 million people. [...]"
        "Query: Does this text mention any physical assets, locations, and ownerships?\n"
        "physical assets: []\nlocations: [Republic of Indonesia, Malaysia, Australia]\nownerships: []\ncommodities: []\n\n"
        "relationships: \n"
\end{lstlisting}
\end{tcolorbox}

\begin{tcolorbox}[breakable, colback=white!5, colframe=black!75, title=Chain-of-Thought (CoT) prompting]
\begin{lstlisting}[basicstyle=\small\ttfamily, breaklines=true]
f"Text: {chunk}\nQuery:  Let's think step by step. First, identify any physical assets mentioned in the text. Next, determine if any locations or ownership details are provided for these physical assets. Then, determine if the commodities related to the physical assets are provided. Finally, summarize the relationships between each physical asset, its location, its ownership and its commodity. "
        "If yes, please specify them in the following format:\n"
        "physical assets: [ ]\nlocations: [ ]\nownerships: [ ]\ncommodities: []\n"
        "Additionally, identify the relationships between them, specifying the location of each physical asset, their ownership details, and commodities. "
        "Format the relationships as follows:\nrelationships: [asset: '', location: '', ownership: '', commodities: '']"
    
\end{lstlisting}
\end{tcolorbox}

\begin{tcolorbox}[breakable, colback=white!5, colframe=black!75, title=Generated knowledge prompting]
\begin{lstlisting}[basicstyle=\small\ttfamily, breaklines=true]
    knowledge_prompt = (
        "You are an expert in analyzing texts for information about physical assets, locations, ownerships, and commodities. "
        "Provide a brief summary of how to identify these elements in a text and the relationships between them."
    )
    prompt_step1 = (
        f"{generated_knowledge}\n\n"
        "You are a virtual assistant with expertise in extracting specific information from text. "
        "A physical asset is an asset with a geographical location.\n\n"
        f"Text: {chunk}\nQuery: Identify any physical assets mentioned in the text. "
        "List them in the format:\nphysical assets: [ ]"
    )
    prompt_step2 = (
        f"{generated_knowledge}\n\n"
        "Using the extracted physical assets:\n"
        f"physical assets: {physical_assets}\n\n"
        f"Text: {chunk}\nQuery: Identify any locations mentioned in the text associated with the physical assets. "
        "List them in the format:\nlocations: [ ]"
    )
    prompt_step3 = (
        f"{generated_knowledge}\n\n"
        "Using the extracted physical assets and locations:\n"
        f"physical assets: {physical_assets}\nlocations: {locations}\n\n"
        f"Text: {chunk}\nQuery: Identify any ownership details mentioned in the text associated with the physical assets. "
        "List them in the format:\nownerships: [ ]"
    )
    prompt_step4 = (
        f"{generated_knowledge}\n\n"
        "Using the extracted physical assets, locations, and ownerships:\n"
        f"physical assets: {physical_assets}\nlocations: {locations}\nownerships: {ownerships}\n\n"
        f"Text: {chunk}\nQuery: Identify any commodities mentioned in the text associated with the physical assets. "
        "List them in the format:\ncommodities: [ ]"
    )    
    prompt_step5 = (
        f"{generated_knowledge}\n\n"
        "Using the extracted physical assets, locations, ownerships, and commodities:\n"
        f"physical assets: {physical_assets}\nlocations: {locations}\nownerships: {ownerships}\ncommodities: {commodities}\n\n"
        "Text: {chunk}\nQuery: Identify the relationships between the physical assets, locations, ownerships, and commodities. "
        "Format the relationships as follows:\nrelationships: [asset: '', location: '', ownership: '', commodities: '']"
    )    
\end{lstlisting}
\end{tcolorbox}

\begin{tcolorbox}[breakable, colback=white!5, colframe=black!75, title=Prompt Chaining]
\begin{lstlisting}[basicstyle=\small\ttfamily, breaklines=true]
    prompt_step1 = (
        f"Text: {chunk}\nQuery: Does this text mention any physical assets? "
        "If yes, please specify them in the following format:\n"
        "physical assets: [ ]"
    )

    prompt_step2 = (
        f"physical assets: {physical_assets}\n"
        f"Text: {chunk}\nQuery: Does this text mention any locations associated with the physical assets? "
        "If yes, please specify them in the following format:\n"
        "locations: [ ]"
    )

    prompt_step3 = (
        f"physical assets: {physical_assets}\nlocations: {locations}\n"
        f"Text: {chunk}\nQuery: Does this text mention any ownership details associated with the physical assets? "
        "If yes, please specify them in the following format:\n"
        "ownerships: [ ]"
    )

    prompt_step4 = (
        f"physical assets: {physical_assets}\nlocations: {locations}\nownerships: {ownerships}\n"
        f"Text: {chunk}\nQuery: Does this text mention any commodities associated with the physical assets? "
        "If yes, please specify them in the following format:\n"
        "commodities: [ ]"
    )

    prompt_step5 = (
        f"physical assets: {physical_assets}\nlocations: {locations}\nownerships: {ownerships}\ncommodities: {commodities}\n"
        f"Text: {chunk}\nQuery: Identify the relationships between the physical assets, locations, ownerships, and commodities. "
        "Format the relationships as follows:\nrelationships: [asset: '', location: '', ownership: '', commodities: '']"
    )
\end{lstlisting}
\end{tcolorbox}

\begin{tcolorbox}[breakable, colback=white!5, colframe=black!75, title=Role prompting]
\begin{lstlisting}[basicstyle=\small\ttfamily, breaklines=true]
    prompt_instruction = (
        "You are a virtual assistant with advanced expertise in a broad spectrum of topics, equipped to utilize high-level critical thinking, cognitive skills, creativity, and innovation.\n"
        "Your goal is to deliver the most straightforward and accurate answer possible for each question, ensuring high-quality and useful responses for the user.\n"
        "Now, let's analyze the following text:\n"
        f"Text: {chunk}\nQuery: Does this text mention any physical assets, locations or ownerships? Does the text mention what commodity the physical asset is being used for?\n"
        "If yes, you must specify them in the following format:\n"
        "physical assets: [ ]\nlocations: [ ]\nownerships: [ ]\ncommodities: []\n"
        "Additionally, identify the relationships between them, specifying the location of each physical asset, the ownership details, the commodity the physical asset is used for and the status of the physical asset. "
        "Format the relationships as follows:\nrelationships: [asset: '', location: '', ownership: '', commodity: '']."
    )
\end{lstlisting}
\end{tcolorbox}

\begin{tcolorbox}[breakable, colback=white!5, colframe=black!75, title=Role + instructional prompting]
\begin{lstlisting}[basicstyle=\small\ttfamily, breaklines=true]
        "You are a virtual assistant with advanced expertise in a broad spectrum of topics, equipped to utilize high-level critical thinking, cognitive skills, creativity, and innovation.\n"
        "Your goal is to deliver the most straightforward and accurate answer possible for each question, ensuring high-quality and useful responses for the user.\n"
        "A physical asset is a tangible resource that a company owns and uses in the production of goods and services. Examples of physical assets are facilities, equipment, infrastructure, etc. Ensure that a geographical location or region is never considered as an asset.\n"
        "A financial asset or other non-physical asset should never be included as a physical asset. Examples of financial assets include equity commitments, corporate facilities, accounts receivable, and short-term investments. Never include these in the list of physical assets.\n"
        "A commodity is what the physical asset is being used for. Examples include copper, gold, electricity, renewable energy, etc."
        "Now, let's analyze the following text:\n"
        f"Text: {chunk}\nQuery: Does this text mention any physical assets, locations or ownerships? Does the text mention what commodity the physical asset is being used for?\n"
        "If yes, you must specify them in the following format:\n"
        "physical assets: [ ]\nlocations: [ ]\nownerships: [ ]\ncommodities: []\n"
        "Additionally, identify the relationships between them, specifying the location of each physical asset, the ownership details, and the commodity the physical asset is used for."
        "Format the relationships as follows:\nrelationships: [asset: '', location: '', ownership: '', commodity: '']. Do not output anything else."
    )
\end{lstlisting}
\end{tcolorbox}

\begin{tcolorbox}[breakable, colback=white!5, colframe=black!75, title=IRZ-CoT prompting]
\begin{lstlisting}[basicstyle=\small\ttfamily, breaklines=true]
    prompt_instruction = (
        "You are a virtual assistant with advanced expertise in a broad spectrum of topics, equipped to utilize high-level critical thinking, cognitive skills, creativity, and innovation.\n"
        "Your goal is to deliver the most straightforward and accurate answer possible for each question, ensuring high-quality and useful responses for the user.\n"
        "A physical asset is a tangible resource that a company owns and uses in the production of goods and services. Examples of physical assets are facilities, equipment, infrastructure, etc. Ensure that a geographical location or region is never considered as an asset.\n"
        "A financial asset or other non-physical asset should never be included as a physical asset. Examples of financial assets include equity commitments, corporate facilities, accounts receivable, and short-term investments. Never include these in the list of physical assets.\n"
        "A commodity is what the physical asset is being used for. Examples include copper, gold, electricity, renewable energy, etc."
        "Now, let's analyze the following text:\n"
        f"Text: {chunk}\nQuery: Let's think step-by-step. Does this text mention any physical assets, locations or ownerships? Does the text mention what commodity the physical asset is being used for?\n"
        "If yes, you must specify them in the following format:\n"
        "physical assets: [ ]\nlocations: [ ]\nownerships: [ ]\ncommodities: []\n"
        "Additionally, identify the relationships between them, specifying the location of each physical asset, the ownership details, and the commodity the physical asset is used for."
        "Format the relationships as follows:\nrelationships: [asset: '', location: '', ownership: '', commodity: '']. Do not output anything else."
    )
\end{lstlisting}
\end{tcolorbox}

\begin{tcolorbox}[breakable, colback=white!5, colframe=black!75, title=Dynamic prompting]
\begin{lstlisting}[basicstyle=\small\ttfamily, breaklines=true]
    prompt_instruction = (
        "You are a virtual assistant with advanced expertise in a broad spectrum of topics, equipped to utilize high-level critical thinking, cognitive skills, creativity, and innovation.\n"
        "Your goal is to deliver the most straightforward and accurate answer possible for each question, ensuring high-quality and useful responses for the user.\n"
    )

    if contains_assets:
        prompt_instruction += (
            "A physical asset is a tangible resource that a company owns and uses in the production of goods and services. Examples of physical assets are facilities, equipment, infrastructure, etc.\n"
            "Ensure that a geographical location or region is never considered as an asset.\n"
            "A financial asset or other non-physical asset should never be included as a physical asset. Examples of financial assets include equity commitments, corporate facilities, accounts receivable, and short-term investments. Never include these in the list of physical assets.\n"
        )

    if contains_commodities:
        prompt_instruction += (
            "A commodity is what the physical asset is being used for. Examples include copper, gold, electricity, renewable energy, etc.\n"
        )

    if contains_locations:
        prompt_instruction += (
            "Always ensure that a geographical location or region is mentioned separately from the physical asset.\n"
        )

    prompt_instruction += (
        f"Now, let's analyze the following text:\n"
        f"Text: {chunk}\nQuery: Let's think step-by-step. Does this text mention any physical assets, locations or ownerships? Does the text mention what commodity the physical asset is being used for?\n"
        "If yes, you must specify them in the following format:\n"
        "physical assets: [ ]\nlocations: [ ]\nownerships: [ ]\ncommodities: []\n"
        "Additionally, identify the relationships between them, specifying the location of each physical asset, the ownership details, and the commodity the physical asset is used for."
        "Format the relationships as follows:\nrelationships: [asset: '', location: '', ownership: '', commodity: '']. Do not output anything else."
    )
\end{lstlisting}
\end{tcolorbox}

\begin{tcolorbox}[breakable, colback=white!5, colframe=black!75, title=Database Cleaning]
\begin{lstlisting}[basicstyle=\small\ttfamily, breaklines=true]
You are an expert data cleaner. Your task is to clean and standardize the following text. You will be provided each cell value one by one with its respective column name. Apply the following cleaning steps:

   - Standardize entries in the commodity column to have a consistent format. For example, "Silver, Gold, Lead, Zinc" should be the standard format for each commodity, separated by commas and no extra spaces.
   - Ensure all entries in the status column are in a consistent format, removing redundant words or phrases.
   - All entries should be in title case.
   - Do not make any changes to the 'Countries' column.
   - In the 'commodity' column, if chemical symbols are given, change these to the element name corresponding to the chemical symbol.
   - In all the columns, ensure each entry is properly formatted without redundant commas and extra spaces. For example, "ExxonMobil" should not be separated by extra commas.
   - Remove any leading or trailing spaces in all columns.
   - All individual commodities should be separated by a comma.
   - The 'location' column should only consist of geographical regions and locations.
   - A physical asset is a tangible resource that a company owns and uses in the production of goods and services. Examples of physical assets are facilities, equipment, infrastructure, etc. If there are any entries in the physical asset column that do not fit the description of a physical asset, put N/A in the corresponding cell.
   - A commodity is what the physical asset is being used for. If there are any entries in the commodity column that do not fit the description of a commodity, put N/A next to the word in brackets.
   - Ensure that there are no repetitions or redundant entries in any of the cells.
   - If any cell has 'not specified', it should be empty.
   - All cells should have standardized entries.

Process the following text according to these instructions. Return only the new cleaned cell value, nothing else.
\end{lstlisting}
\end{tcolorbox}

\begin{tcolorbox}[breakable, colback=white!5, colframe=black!75, title=Country Extraction Prompt]
\begin{lstlisting}[basicstyle=\ttfamily, breaklines=true]
You are an expert in geographical locations. Given the location information provided, identify the countries mentioned in the location. Return the list of countries separated by commas. If no country is mentioned, return "N/A".
\end{lstlisting}
\end{tcolorbox}

\begin{tcolorbox}[breakable, colback=white!5, colframe=black!75, title=Modified prompt for improvement module]
\begin{lstlisting}[basicstyle=\small\ttfamily, breaklines=true]
"You are a virtual assistant with advanced expertise in a broad spectrum of topics, equipped to utilize high-level critical thinking, cognitive skills, creativity, and innovation.\n"
        "Your goal is to deliver the most straightforward and accurate answer possible for each question, ensuring high-quality and useful responses for the user.\n"
        "A physical asset is a tangible resource that a company owns in a location and uses in the production of goods and services. Examples of physical assets are all examples of 'Plant' in the tables (Wateree, Greensville and Colonial Trail West are all physical assets).\n"
        "A financial asset or other non-physical asset should never be included as a physical asset. Examples of financial assets include equity commitments, corporate facilities, accounts receivable, and short-term investments. Never include these in the list of physical assets.\n"
        "A commodity is what the physical asset is being used for. The status of a physical asset gives information on whether the asset is operational, under construction or in end-of-life."
        "Now, let's analyze the following text:\n"
        f"Text: {text}\nQuery: Let's think step-by-step. Does this text mention any physical assets, locations or ownerships? Does the text mention what commodity the physical asset is being used for?\n"
        "Does the text mention the status of the physical asset? Examples of status include whether the asset is operational, under construction or in end-of-life."
        "If yes, you must specify them in the following format:\n"
        "physical assets: [ ]\nlocations: [ ]\nownerships: [ ]\ncommodities: []\nstatus: []\n"
        "Additionally, identify all the relationships between them, specifying the location of each physical asset, the ownership details, the commodity the physical asset is used for and the status of the physical asset. Do not leave out any relationships. "
        "Format the relationships as follows:\nrelationships: [asset: '', location: '', ownership: '', commodity: '', status: '']. Do not output anything else."
\end{lstlisting}
\end{tcolorbox}

\subsection{Standardised Characters in Foundational Data Cleaning \& Standardisation}
In this section, we detail the standardisation rules applied during the Foundational Data Cleaning \& Standardisation phase in our database cleaning process. By enforcing these cleaning and standardisation rules, the data becomes more reliable and easier to analyse, reducing potential errors caused by inconsistent naming and formatting practices. This standardisation forms the foundation for subsequent cleaning steps within the study.

\begin{itemize}
    \item \textbf{Characters Standardised:}
    \begin{itemize}
        \item \textbackslash\ (Backslashes): Removed from text fields.
        \item \textquotesingle\ (Single quotes): Removed from text fields.
        \item '' (Double quotes): Removed from text fields.
        \item \textbf{Extra spaces:}
        \begin{itemize}
            \item Leading and trailing spaces: Trimmed from text entries.
            \item Multiple consecutive spaces: Condensed to a single space.
        \end{itemize}
        \item \textbf{Commas (,)}: Ensured proper spacing after commas by replacing them with \texttt{`,\ '} (comma followed by a space).
    \end{itemize}

    \item \textbf{Names Standardised:}
    \begin{itemize}
        \item \textbf{Company Names in \texttt{ownership} Field:}
        \begin{itemize}
            \item All the following aliases and variations are standardized to \textbf{``Newmont Corporation''}:
            \begin{itemize}
                \item ``Company''
                \item ``company''
                \item ``The company''
                \item ``the company''
                \item ``The Company''
                \item ``Company's''
                \item ``the Company''
                \item ``we''
                \item ``NEWMONT CORPORATION''
                \item ``Newmont's ownership or economic interest''
                \item ``Company owns or controls land''
                \item ``Newmont''
                \item ``Newmont (majority)''
                \item ``Newmont Corporation (formerly)''
                \item ``100\% owned by the Company''
                \item ``Newmont Stockholders''
                \item ``100\% by Newmont''
                \item ``Company owned''
                \item ``Company's''
            \end{itemize}
        \end{itemize}
        \item \textbf{USA Variants in \texttt{location} Field:}
        \begin{itemize}
            \item All the following variants are standardised to \textbf{``USA''}:
            \begin{itemize}
                \item ``United States of America''
                \item ``United States''
                \item ``USA''
                \item ``US''
                \item ``USAA''
                \item ``USAA.''
                \item ``U.S.''
                \item ``U.S.A.''
            \end{itemize}
        \end{itemize}
    \end{itemize}
\end{itemize}

\subsection{LSEG Database Numerical Results}
In this section, we present the complete numerical results from stage 1 of the validation, showing six performance scores for each company.

\clearpage
\onecolumn
\begin{scriptsize}
\renewcommand{\arraystretch}{1.1}
\begin{longtable}{|p{1.0cm}|p{1.0cm}|p{1.8cm}|p{1.2cm}|p{1.2cm}|p{1.2cm}|p{1.2cm}|p{1.2cm}|p{1.2cm}|}
\caption{Complete Performance Scores for Companies across the following metrics: Partial Match Score, Jaccard Similarity, Cosine Similarity, Dice Similarity Coefficient, Normalised Levenshtein Distance, and Hits@5.}
\label{tab:companymetrics} \\ \hline
\multicolumn{1}{|c|}{\textbf{Sector}} & 
\multicolumn{1}{c|}{\textbf{Company}} & 
\multicolumn{1}{c|}{\textbf{Entity}} & 
\multicolumn{1}{c|}{\makecell{\textbf{Partial} \\ \textbf{Match}}} & 
\multicolumn{1}{c|}{\makecell{\textbf{Jaccard} \\ \textbf{Similarity}}} & 
\multicolumn{1}{c|}{\makecell{\textbf{Cosine} \\ \textbf{Similarity}}} & 
\multicolumn{1}{c|}{\makecell{\textbf{Dice} \\ \textbf{Similarity}}} & 
\multicolumn{1}{c|}{\makecell{\textbf{Norm.} \\ \textbf{Levenshtein}}} & 
\multicolumn{1}{c|}{\makecell{\textbf{Hits@5}}}\\ \hline
\endfirsthead

\hline
\multicolumn{9}{|c|}{\textbf{(continued)}} \\ \hline
\multicolumn{1}{|c|}{\textbf{Sector}} & 
\multicolumn{1}{c|}{\textbf{Company}} & 
\multicolumn{1}{c|}{\textbf{Entity}} & 
\multicolumn{1}{c|}{\makecell{\textbf{Partial} \\ \textbf{Match}}} & 
\multicolumn{1}{c|}{\makecell{\textbf{Jaccard} \\ \textbf{Similarity}}} & 
\multicolumn{1}{c|}{\makecell{\textbf{Cosine} \\ \textbf{Similarity}}} & 
\multicolumn{1}{c|}{\makecell{\textbf{Dice} \\ \textbf{Similarity}}} & 
\multicolumn{1}{c|}{\makecell{\textbf{Norm.} \\ \textbf{Levenshtein}}} & 
\multicolumn{1}{c|}{\makecell{\textbf{Hits@5}}}\\ \hline
\endhead

\hline \multicolumn{9}{|r|}{{Continued on next page}} \\ \hline
\endfoot

\hline
\endlastfoot

\multirow{4}{*}{Mining} & \multirow{4}{*}{AA} 
& Physical Asset & 0.95 & 0.7833 & 0.8591 & 0.8333 & 0.7667 & 0.7143 \\ \cline{3-9}
& & Ownership    & 0.93 & 0.2139 & 0.3609 & 0.3389 & 0.3658 & 0.7143 \\ \cline{3-9}
& & Commodity    & 0.52 & 0.0000 & 0.0000 & 0.0000 & 0.1839 & 0.7143 \\ \cline{3-9}
& & Countries    & 0.100 & 0.8333 & 0.8963 & 0.8333 & 0.8639 & 0.7143 \\ \cline{2-9}

& \multirow{4}{*}{SCCO} 
& Physical Asset & 0.81 & 0.6389 & 0.7416 & 0.7333 & 0.6349 & 0.5455 \\ \cline{3-9}
& & Ownership    & 1.00 & 0.3000 & 0.5375 & 0.4127 & 0.5590 & 0.5455 \\ \cline{3-9}
& & Commodity    & 0.68 & 0.0000 & 0.5244 & 0.0000 & 0.1966 & 0.5455 \\ \cline{3-9}
& & Countries    & 0.82 & 0.6667 & 0.6667 & 0.6667 & 0.7222 & 0.5455 \\ \cline{2-9}

& \multirow{4}{*}{FCX} 
& Physical Asset & 0.88 & 0.5667 & 0.7014 & 0.6672 & 0.5148 & 0.8462 \\ \cline{3-9}
& & Ownership    & 0.93 & 0.1667 & 0.7063 & 0.2222 & 0.5020 & 0.8462 \\ \cline{3-9}
& & Commodity    & 0.87 & 0.0000 & 0.3668 & 0.0000 & 0.2114 & 0.7692 \\ \cline{3-9}
& & Countries    & 1.00 & 1.0000 & 1.0000 & 1.0000 & 1.0000 & 0.8462 \\ \cline{2-9}

& \multirow{4}{*}{NEM} 
& Physical Asset & 0.83 & 0.8462 & 0.9083 & 0.8949 & 0.8531 & 0.6486 \\ \cline{3-9}
& & Ownership    & 0.89 & 0.6520 & 0.7891 & 0.7090 & 0.7747 & 0.6216 \\ \cline{3-9}
& & Commodity    & 0.83 & 0.1154 & 0.5488 & 0.1282 & 0.3698 & 0.5946 \\ \cline{3-9}
& & Countries    & 0.79 & 0.7692 & 0.7692 & 0.7692 & 0.8225 & 0.6486 \\ \cline{2-9}

& \multirow{4}{*}{HL} 
& Physical Asset & 0.82 & 0.7955 & 0.8608 & 0.8515 & 0.7738 & 0.6667 \\ \cline{3-9}
& & Ownership    & 0.97 & 0.5000 & 0.6667 & 0.6667 & 0.7500 & 0.6667 \\ \cline{3-9}
& & Commodity    & 0.88 & 0.0000 & 0.5776 & 0.0000 & 0.3281 & 0.6111 \\ \cline{3-9}
& & Countries    & 0.62 & 0.1818 & 0.1818 & 0.1818 & 0.3566 & 0.6667 \\ \hline

\multirow{16}{*}{Oil \& Gas} 
& \multirow{4}{*}{CVX} 
& Physical Asset & 0.72 & 0.3417 & 0.5084 & 0.5016 & 0.4292 & 0.6250 \\ \cline{3-9}
& & Ownership    & 1.00 & 1.0000 & 1.0000 & 1.0000 & 1.0000 & 0.6250 \\ \cline{3-9}
& & Commodity    & 0.36 & 0.0000 & 0.4472 & 0.0000 & 0.2857 & 0.1250 \\ \cline{3-9}
& & Countries    & 0.53 & 0.0000 & 0.0000 & 0.0000 & 0.2000 & 0.5000 \\ \cline{2-9}

& \multirow{4}{*}{XOM} 
& Physical Asset & 0.71 & 0.2418 & 0.4341 & 0.3840 & 0.3808 & 0.6364 \\ \cline{3-9}
& & Ownership    & 0.94 & 0.4286 & 0.5469 & 0.5238 & 0.5276 & 0.6364 \\ \cline{3-9}
& & Commodity    & 0.35 & 0.0000 & 0.0000 & 0.0000 & 0.1852 & 0.0000 \\ \cline{3-9}
& & Countries    & 0.63 & 0.1429 & 0.1667 & 0.1429 & 0.3127 & 0.5455 \\ \cline{2-9}

& \multirow{4}{*}{MPC} 
& Physical Asset & 0.74 & 0.3026 & 0.4935 & 0.4556 & 0.3552 & 0.7895 \\ \cline{3-9}
& & Ownership    & 0.95 & 0.6083 & 0.7457 & 0.6925 & 0.6947 & 0.7368 \\ \cline{3-9}
& & Commodity    & 0.49 & 0.1036 & 0.4149 & 0.1693 & 0.1907 & 0.5789 \\ \cline{3-9}
& & Countries    & 0.43 & 0.4722 & 0.6855 & 0.6389 & 0.2493 & 0.7895 \\ \cline{2-9}

& \multirow{4}{*}{COP} 
& Physical Asset & 0.76 & 0.2000 & 0.3536 & 0.3333 & 0.2500 & 1.0 \\ \cline{3-9}
& & Ownership    & 1.00 & 0.3333 & 0.5774 & 0.5000 & 0.4118 & 1.0 \\ \cline{3-9}
& & Commodity    & 1.00 & 0.3333 & 0.8944 & 0.5000 & 0.1667 & 1.0 \\ \cline{3-9}
& & Countries    & 0.67 & 0.5000 & 0.7071 & 0.6667 & 0.2727 & 1.0 \\ \hline

\multirow{16}{*}{Utilities} 
& \multirow{4}{*}{D} 
& Physical Asset & 0.79 & 0.2532 & 0.6368 & 0.3961 & 0.3788 & 0.2778 \\ \cline{3-9}
& & Ownership    & 0.85 & 0.2740 & 0.4884 & 0.4015 & 0.4966 & 0.3333 \\ \cline{3-9}
& & Commodity    & 0.75 & 0.2292 & 0.3836 & 0.3125 & 0.2228 & 0.1667 \\ \cline{3-9}
& & Countries    & 1.00 & 0.9375 & 0.9634 & 0.9583 & 0.8984 & 0.2778 \\ \cline{2-9}

& \multirow{4}{*}{ED} 
& Physical Asset & 0.64 & 0.3500 & 0.6400 & 0.5000 & 0.4602 & 0.5000 \\ \cline{3-9}
& & Ownership    & 1.00 & 0.5000 & 0.7071 & 0.6667 & 0.4637 & 0.5000 \\ \cline{3-9}
& & Commodity    & 0.40 & 0.0000 & 0.0000 & 0.0000 & 0.1364 & 0.0000 \\ \cline{3-9}
& & Countries    & 0.67 & 0.7500 & 0.8500 & 0.8333 & 0.6100 & 0.5000 \\ \cline{2-9}

& \multirow{4}{*}{DUK} 
& Physical Asset & 0.74 & 0.3131 & 0.5573 & 0.4683 & 0.5001 & 0.6000 \\ \cline{3-9}
& & Ownership    & 0.98 & 0.4667 & 0.7953 & 0.6286 & 0.5889 & 0.6000 \\ \cline{3-9}
& & Commodity    & 0.41 & 0.0000 & 0.0000 & 0.0000 & 0.1051 & 0.0000 \\ \cline{3-9}
& & Countries    & 1.00 & 0.5417 & 0.7155 & 0.6667 & 0.3789 & 0.6000 \\ \cline{2-9}

& \multirow{4}{*}{EXC} 
& Physical Asset & 0.93 & 0.3778 & 0.7145 & 0.5367 & 0.5472 & 0.7333 \\ \cline{3-9}
& & Ownership    & 0.71 & 0.2867 & 0.4354 & 0.4333 & 0.4136 & 0.6667 \\ \cline{3-9}
& & Commodity    & 0.40 & 0.4000 & 0.6285 & 0.5333 & 0.3520 & 0.5333 \\ \cline{3-9}
& & Countries    & 0.49 & 0.9000 & 1.0000 & 0.9000 & 1.0000 & 0.6667 \\ \hline

\end{longtable}
\end{scriptsize}
\twocolumn

\subsection{Dashboard User Interface (UI)}
In this section, we present snippets of dashboard user interface (UI) for Alcoa Corporation (AA) as an example of the functionality and design of the visualisation tool developed for this project. The dashboard provides an intuitive and interactive platform to explore the relationships between key entities within the database. 
\begin{figure}[H]
    \centering
    \includegraphics[width=0.9\linewidth]{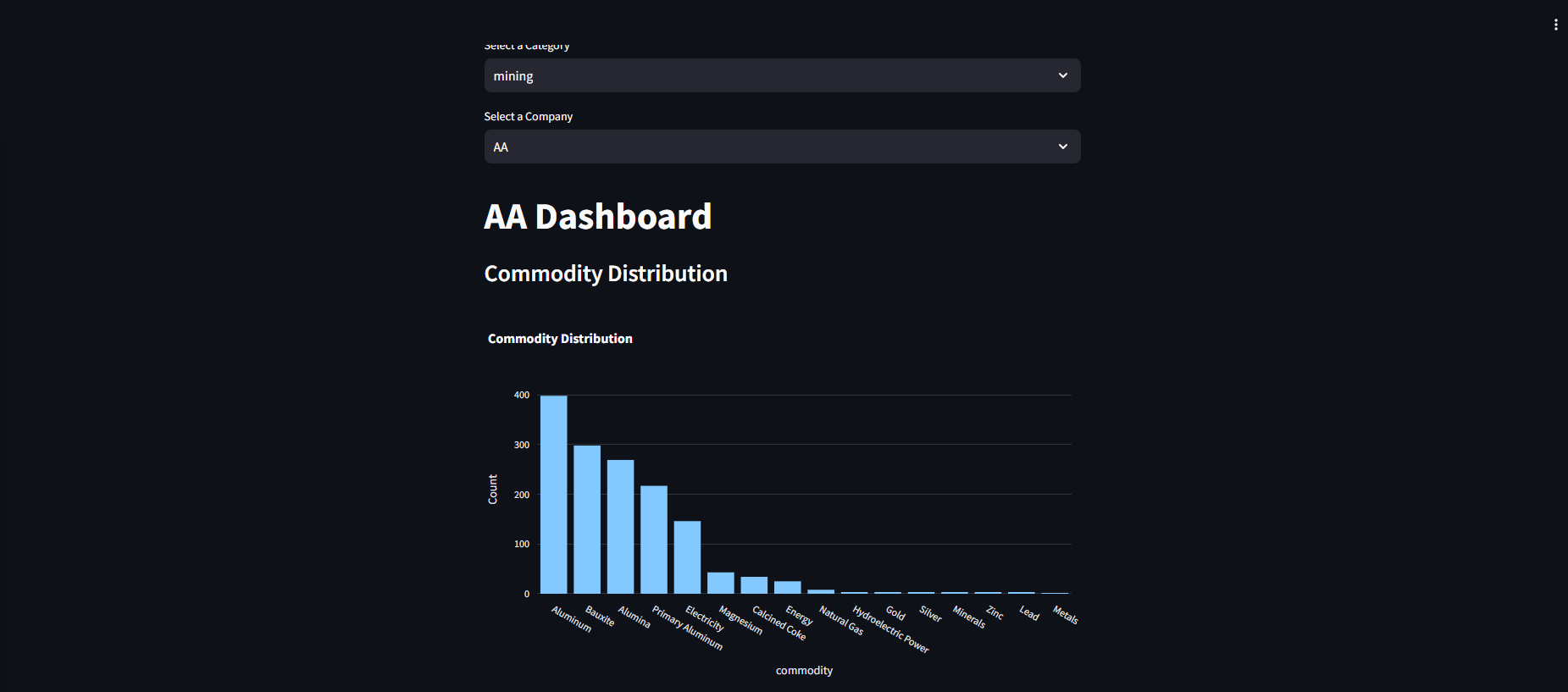}
    \includegraphics[width=0.9\linewidth]{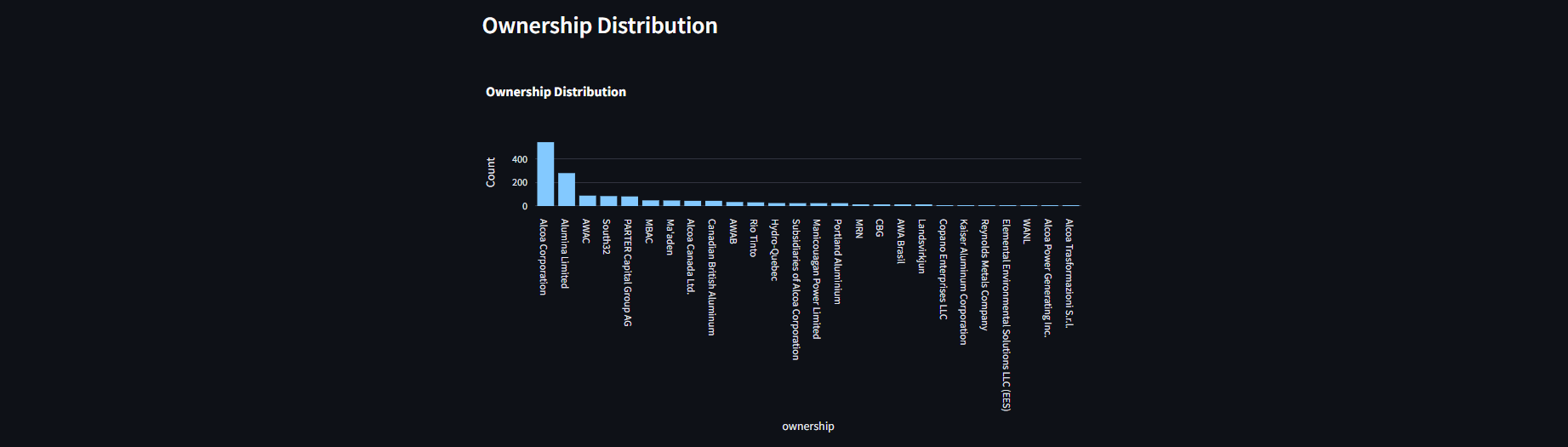}
    \includegraphics[width=0.9\linewidth]{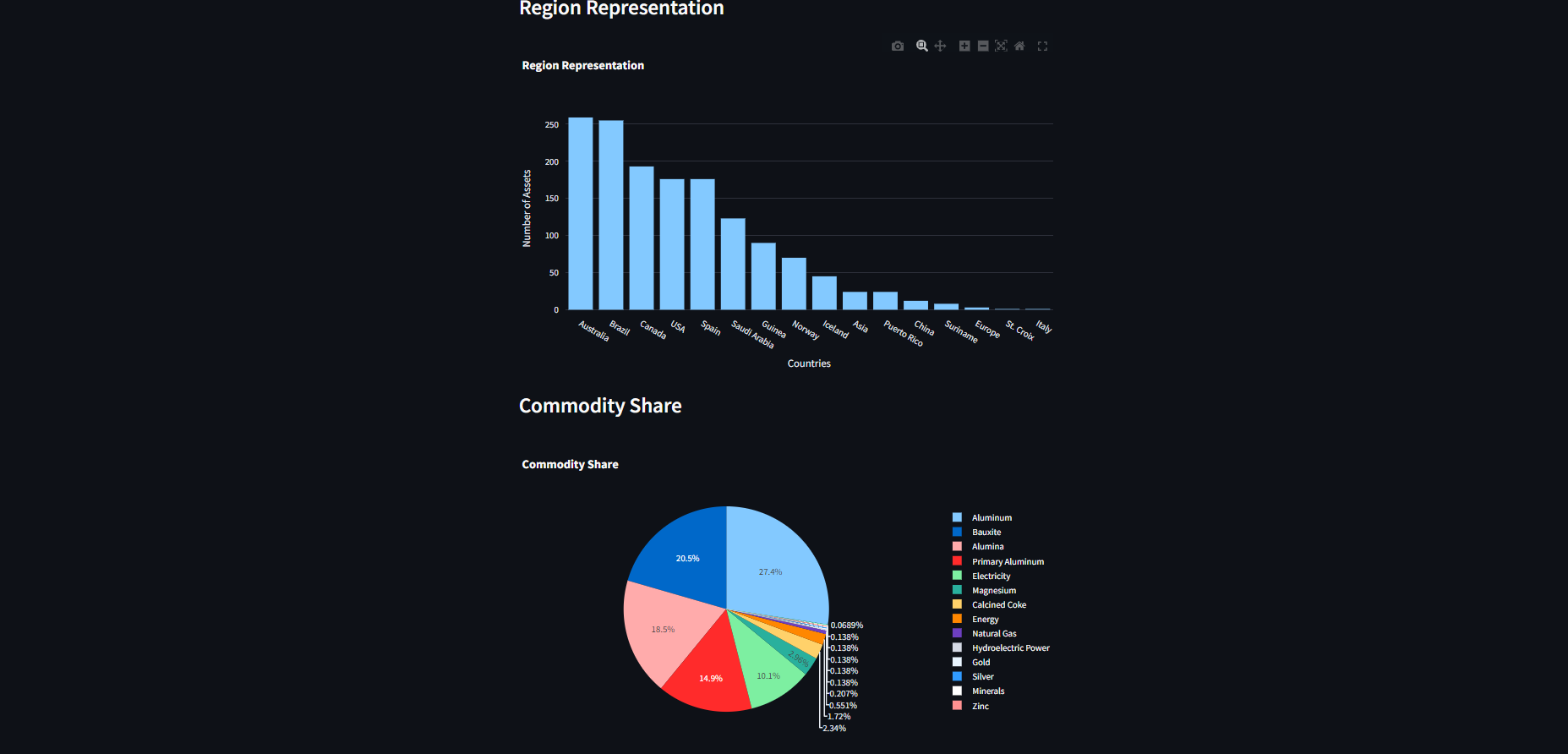}
    \label{fig:dashboards1}
\end{figure}

\begin{figure}[H]
    \centering
    \includegraphics[width=0.9\linewidth]{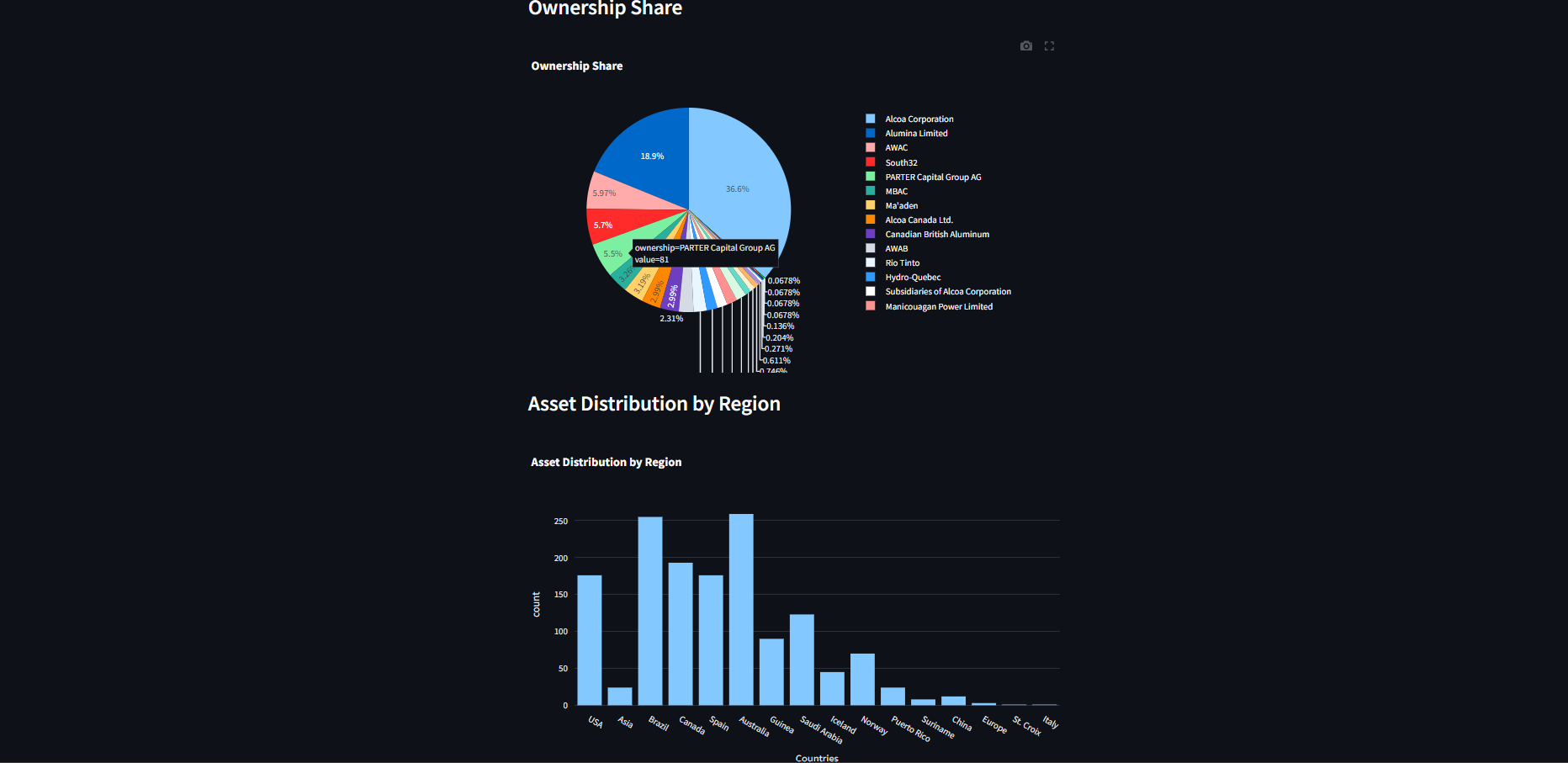}
    \includegraphics[width=0.9\linewidth]{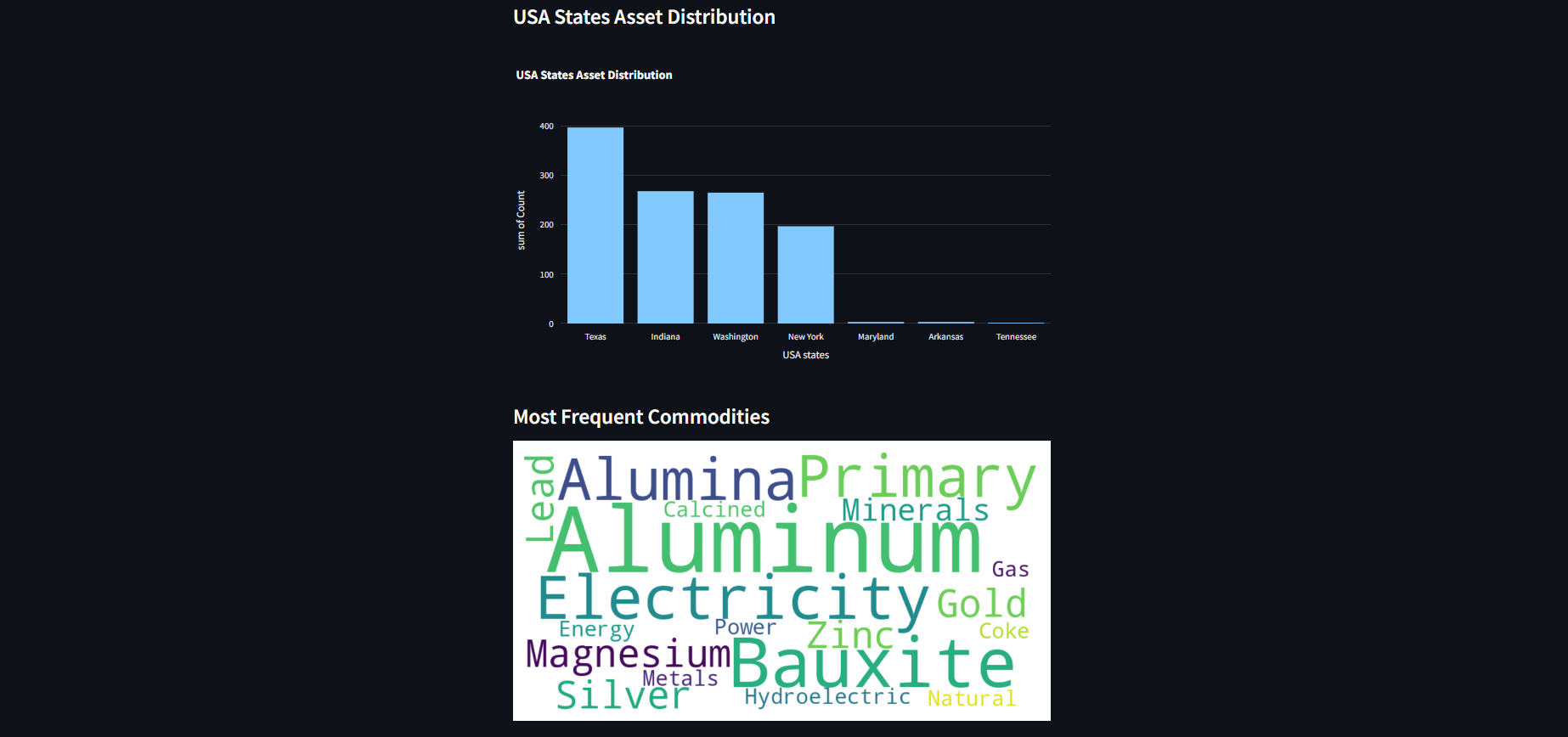}
    \includegraphics[width=0.9\linewidth]{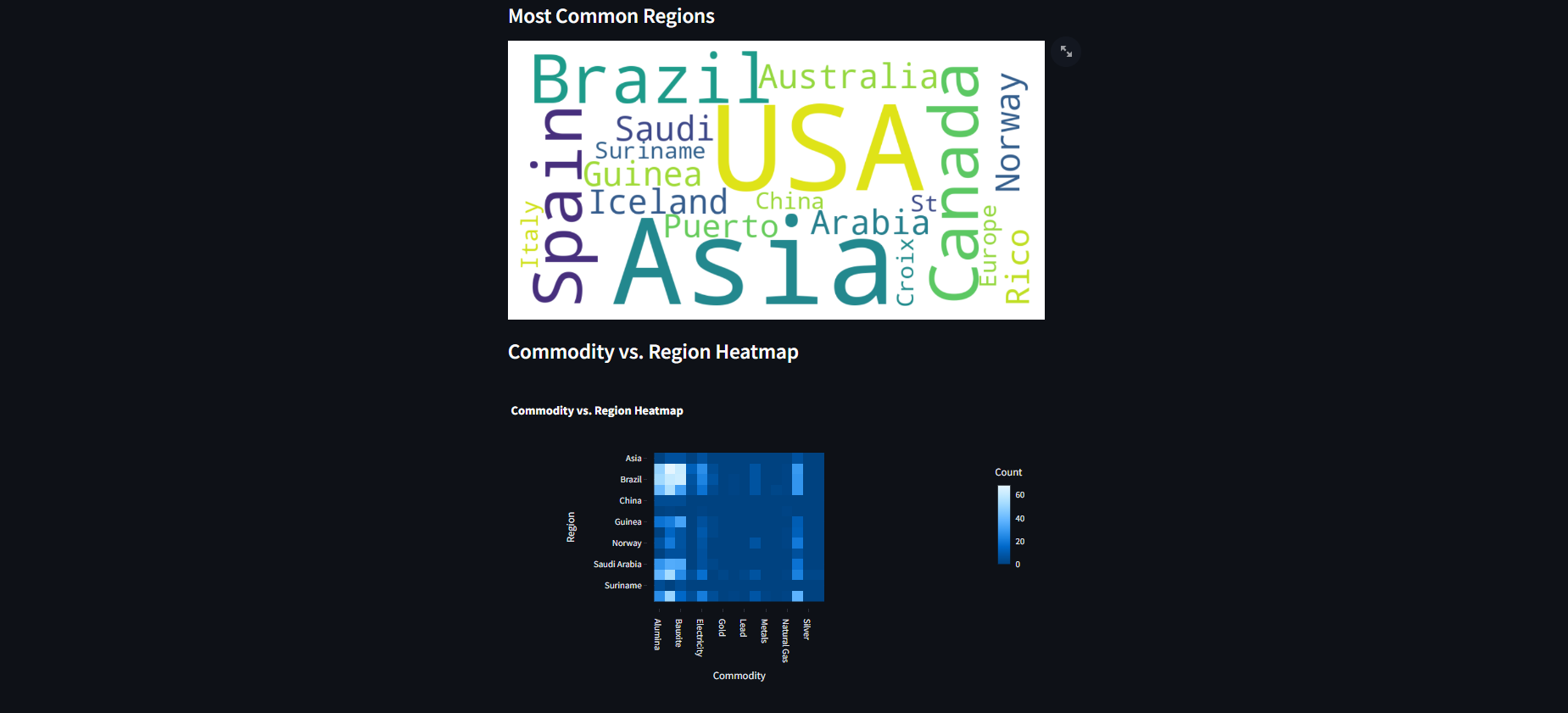}
    \label{fig:dashboards2}
\end{figure}

\begin{figure}[H]
    \centering
    \includegraphics[width=0.9\linewidth]{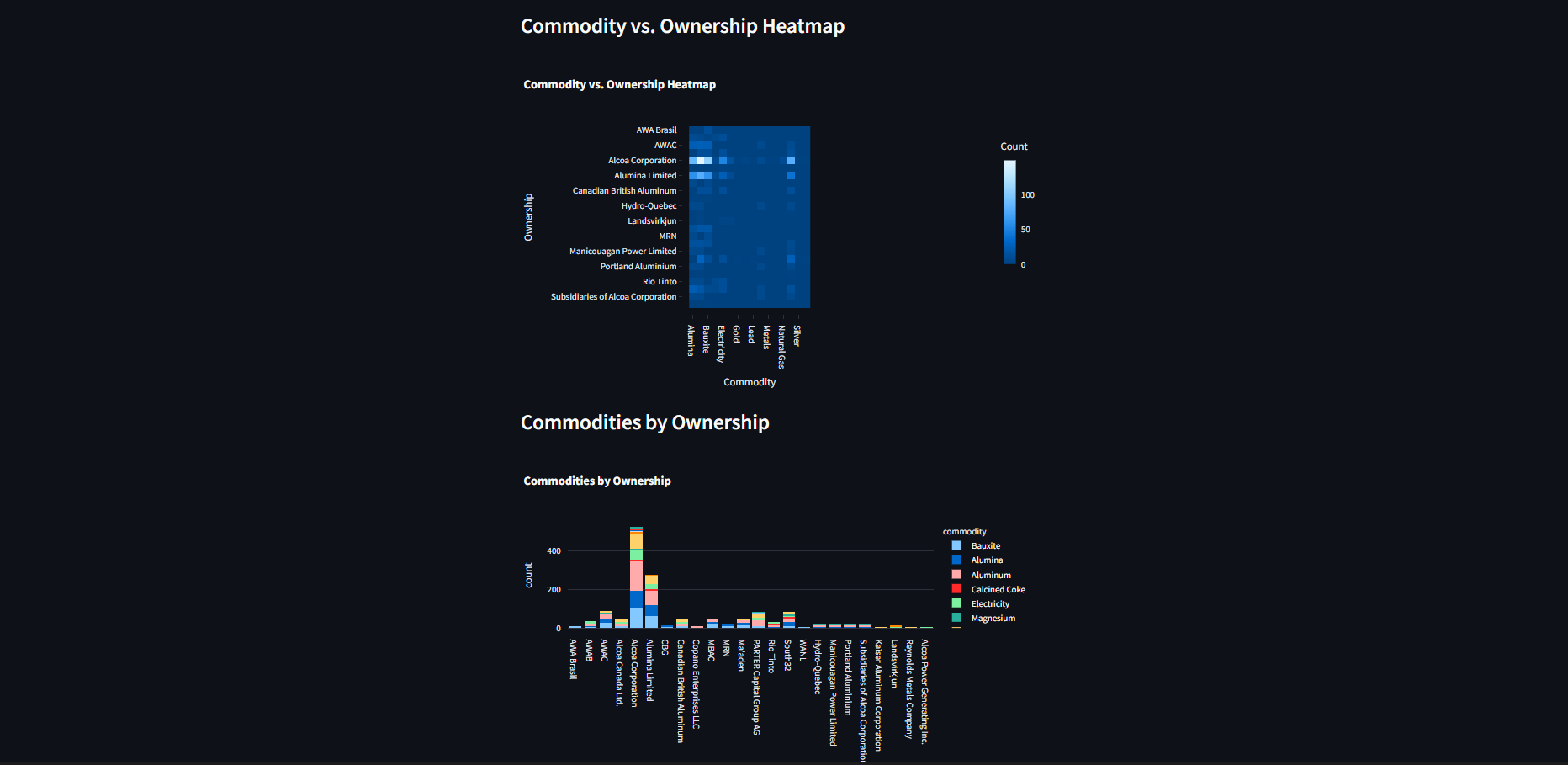}
    \includegraphics[width=0.9\linewidth]{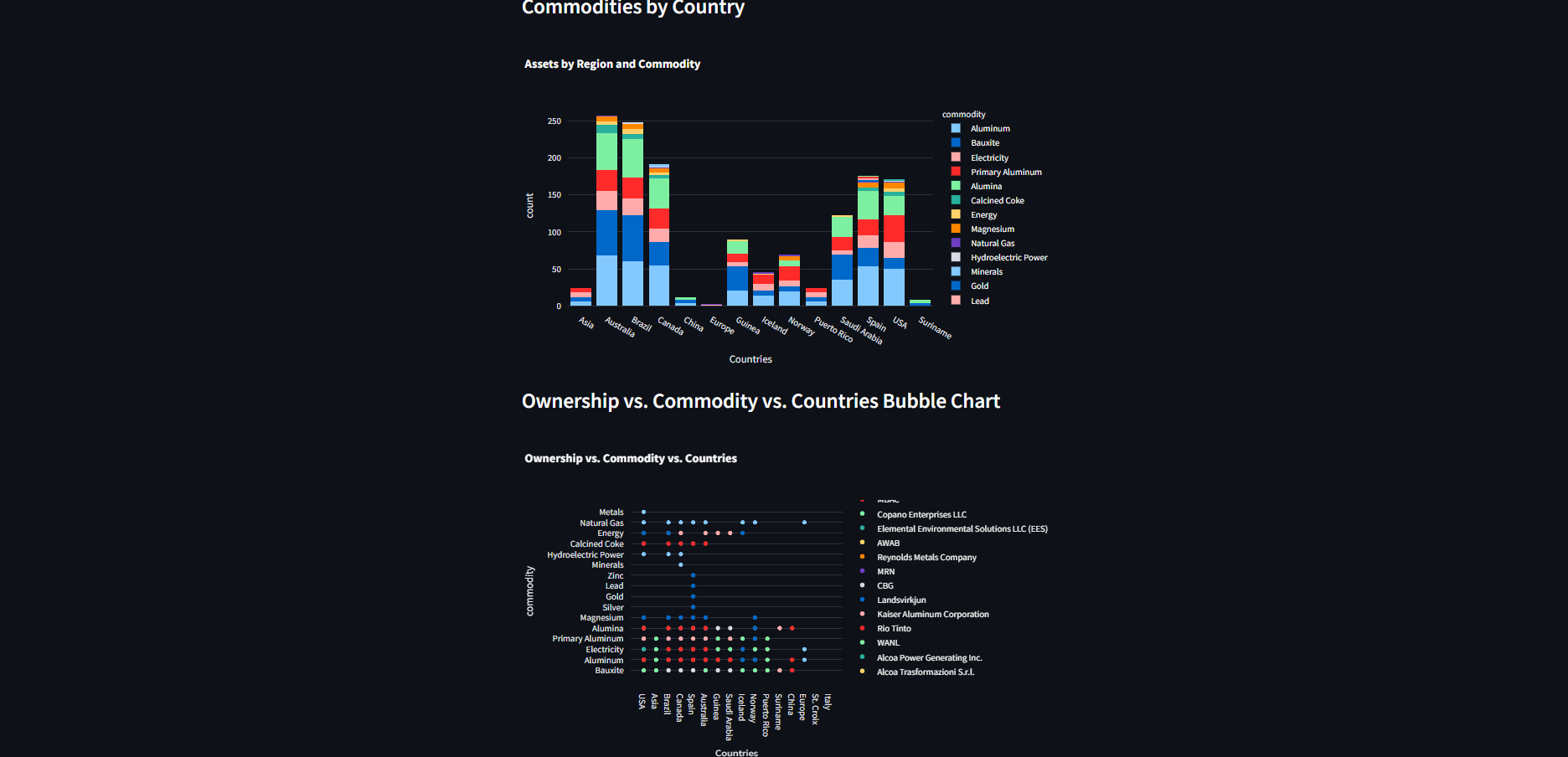}
    \includegraphics[width=0.9\linewidth]{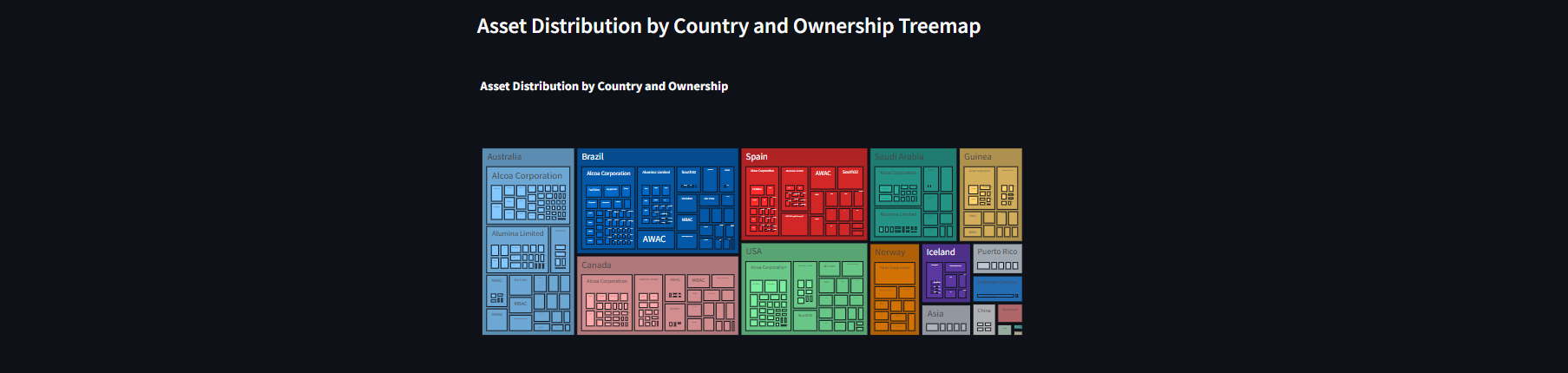}
\end{figure}

\end{document}